\documentclass{aa}
\usepackage{graphicx}
\usepackage{amsmath}
\usepackage{txfonts}
\usepackage{lscape}
\usepackage{longtable, lscape}
\begin{document}
\def\HII{H\,{\sc{ii}}}
\def\SII{S\,{\sc{ii}}}
\def\HI{H\,{\sc{i}}}

\def\Ks{$K_{\rm{s}}$}
\def\h{\hbox{$^{\reset@font\r@mn{h}}$}}
\def\m{\hbox{$^{\reset@font\r@mn{m}}$}}
\def\s{\hbox{$^{\reset@font\r@mn{s}}$}}
\def\msol{\hbox{\kern 0.20em $M_\odot$}}

\def\kms{\hbox{\kern 0.20em km\kern 0.20em s$^{-1}$}}
\def\cmmt{\hbox{\kern 0.20em cm$^{-3}$}}
\def\cmmd{\hbox{\kern 0.20em cm$^{-2}$}}
\def\pc{\hbox{\kern 0.20em pc$^{2}$}}
\def\twco{\hbox{${}^{12}$CO}}
\def\twcotwo{\hbox{${}^{12}$CO(2-1)}}
\def\thco{\hbox{${}^{13}$CO}}
\def\thcotwo{\hbox{${}^{13}$CO(2-1)}}
\def\thcoone{\hbox{${}^{13}$CO(1-0)}}
\def\ceio{\hbox{C${}^{18}$O}}
\def\ceiotwo{\hbox{C${}^{18}$O(2-1)}}
\def\ceioone{\hbox{C${}^{18}$O(1-0)}}
\def\cs{\hbox{CS}}
\def\csthree{\hbox{CS(3-2)}}
\def\cstwo{\hbox{CS(2-1)}}
\def\csfive{\hbox{CS(5-4)}}
\def\cts{\hbox{C${}^{34}$S}}
\def\ctsthree{\hbox{C${}^{34}$S(3-2)}}
\def\ctstwo{\hbox{C${}^{34}$S(2-1)}}
\def\htwo{\hbox{H${}_2$}}
\def\h13cop{\hbox{H$^{13}$CO$^{+}$}}
\def\halpha{\hbox{H$\alpha$ }}
\def\hcop{\hbox{HCO$^{+}$}}
\newcommand{\jonetozero}{\hbox{$J=1\rightarrow 0$}}
\newcommand{\jtwotoone}{\hbox{$J=2\rightarrow 1$}}
\newcommand{\jthreetotwo}{\hbox{$J=3\rightarrow 2$}}
\newcommand{\jfourtothree}{\hbox{$J=4\rightarrow 3$}}
\newcommand{\jfivetofour}{\hbox{$J=5\rightarrow 4$}}
%
   \title{Star formation around RCW~120,\\the perfect
   bubble \thanks{Table~1 is available in electronic form at the CDS 
via anonymous ftp to cdsarc.u-strasbg.fr (130.79.128.5) or via 
http://cdsweb.u-strasbg.fr/cgi-bin/qcat?J/A+A/}
}
%
%
\author{L. Deharveng\inst{1}
\and
A. Zavagno\inst{1}
\and
F. Schuller\inst{2}
\and
J.~Caplan\inst{1}
\and
M.~Pomar\`es\inst{1}
\and 
C. De Breuck\inst{3}
}

   \offprints{L. Deharveng}

   \institute{Laboratoire d'Astrophysique de Marseille (UMR~6110 CNRS \& Universit\'e de Provence),
   38~rue F. Joliot-Curie, 13388 Marseille Cedex 13, France, 
         \and 
         Max-Planck-Institut f{\"u}r Radioastronomie, Auf dem H{\"u}gel 69, 53121 Bonn, Germany,
         \and
         European Southern Observatory, Karl-Schwarschild Strasse, 85748 Garching bei M\"unchen, Germany\\
    \email{lise.deharveng@oamp.fr} 
    }

   \date{Received 12/11/2008; accepted 15/01/2009 }


  \abstract
   { This study deals with the star formation triggered by \HII\ regions.}
   {We wish to take advantage of the very simple morphology of RCW~120~-- 
   a perfect bubble~-- to understand the mechanisms triggering star 
   formation around an \HII\ region and to establish what kind of 
   stars are formed there.}
   {We present 870~$\mu$m observations of RCW~120, 
   obtained with the APEX-LABOCA camera. These show the distribution of cold dust, 
   and thus of neutral material. We use  
   Spitzer-MIPS observations at 24~$\mu$m and 70~$\mu$m to detect the 
   young stellar objects present in this region and to estimate their 
   evolutionary stages.}
   {A layer of dense neutral material surrounds the entire 
   \HII\ region, having been swept up during 
   the region's expansion. This layer has a mass greater than 
   $2000~M_{\sun}$ and is fragmented, with massive fragments 
   elongated along the ionization front (IF). We measured the 24~$\mu$m 
   flux of 138 sources. Of these, 39 are Class~I or flat-spectrum 
   young stellar objects (YSOs) observed 
   in the direction of the collected layer. We show that several
   triggering mechanisms are acting simultaneously in the swept-up 
   shell, where they form a second generation of stars. No massive YSOs 
   are detected. However, a massive, compact 870~$\mu$m 
   core lies adjacent to the IF.\ A 70~$\mu$m source with no 24~$\mu$m 
   counterpart is detected at the same
   position. This source is a likely candidate for a Class~0 YSO.
   Also at 24~$\mu$m, we detect a chain of about ten regularly 
   spaced Class~I or flat spectrum sources, parallel to the IF, 
   in the direction of the most massive fragment. We suggest that the formation 
   of these YSOs is the result of Jeans gravitational instabilities in the 
   collected layer. Finally, the 870~$\mu$m emission, the 24~$\mu$m emission,  
   and the H$\alpha$ emission show the existence of an extended and partially 
   ionized photodissociation region around RCW~120. This demonstrates 
   the long-distance influence of the \HII\ region upon its 
   surrounding medium.}
  {}



\maketitle


\section{Introduction}

RCW~120 is perhaps the most perfect of the many bubbles
detected in the Galactic plane by Spitzer at 8.0~$\mu$m
(Churchwell \cite{chu06}, bubble S7). This \HII\ region presents
a simple morphology, with a well defined ionization front (IF)
separating the ionized and the surrounding neutral material. This
optical \HII\ region has the additional advantage of lying
nearby, at a distance of 1.34~kpc, and being isolated. 
Its distance is reliable as the photometric and kinematic 
determinations are in good agreement (Zavagno et al.~\cite{zav07}; 
hereafter ZA07). The J2000 coordinates of its O8 central exciting star are 
$l =348\fdg2239$, $b = +0\fdg4643$, $\alpha = 17^{\rm h} 12^{\rm m} 20\fs6$,
$\delta = -38\degr 29\arcmin 26\arcsec$. This star is identified 
in ZA07's fig.~1. 

Star formation associated with RCW~120 has been studied by
ZA07, mainly using observations obtained at 1.2~mm, supplemented 
by 2MASS and Spitzer-GLIMPSE data. ZA07 have shown the
presence of a layer of collected neutral material surrounding
the ionized gas. This layer is fragmented, with
massive fragments adjacent to the ionized gas, and elongated along
the IF. Many young stellar
objects (YSOs) are present in the vicinity of the \HII\ region.
However no massive object is observed, even in the direction of the most
massive fragment. ZA07's conclusion is that the collect \& collapse
process is taking place around RCW~120, but that this region is
too young for massive-star formation to take place via
gravitational instabilities along the collected layer.

Since then, new observations have become available, allowing us to progress
in our understanding of star formation triggered by RCW~120. These
observations are: {\it i)} a deep map of the dust emission at
870~$\mu$m, obtained with the APEX-LABOCA camera during the 
science verification phase, giving improved knowledge 
of the cold dust distribution and hence
of the neutral material associated with RCW~120; {\it ii)}
Spitzer-MIPSGAL
frames at 24~$\mu$m and 70~$\mu$m, allowing a better
detection of the YSOs possibly associated with RCW~120, and a better
determination of their evolutionary stages. The 870~$\mu$m
observation is presented in Sect.~2.1, and the 24~$\mu$m and 70~$\mu$m
frames and their photometric reduction in Sect.~2.2. Sect~3 deals
with the distribution of the neutral material around RCW120.  
The nature of the YSOs associated with RCW~120 is discussed
in Sect.~4. In Sect.~5 we try to understand the interactions between
the central \HII\ region and the surrounding material, and how
these interactions influence the star formation in this region.

Additional details about RCW~120, and various composite colour images,  
can be found in Deharveng \& Zavagno~(\cite{deh08}).

\section{New observations}

\subsection{APEX-LABOCA observations at 870~$\mu$m}

The APEX\footnote{This publication is based on data acquired with 
the Atacama Pathfinder EXperiment (APEX), in run 078.F-9005(A). 
APEX is a collaboration between the Max-Planck-Institut 
f\"ur Radioastronomie, the European Southern Observatory, and the 
Onsala Space Observatory.} LABOCA camera was used 
at 870~$\mu$m (345~GHz) to observe the
continuum emission of the cold dust associated with RCW~120. The
Large Apex BOlometer CAmera, commissioned in May 2007,
is a 295-pixel bolometer array developed 
by the Max-Planck-Institut f\"ur Radioastronomie  
(Siringo et al.~\cite{sir07}). The APEX beam size at 870~$\mu$m 
is 19\farcs2. 

RCW~120 was observed in July 2007 as part of the science verification
program. A $15\arcmin\times15\arcmin$ map was obtained in the
spiral raster mode. The on-source observing time was less than two hours.
During these observations the amount of precipitable water 
vapour was $\sim1$~mm. 

The observations were reduced using the Bolometer array data
Analysis package (BoA; Schuller et al.\ in preparation). The reduction 
steps involved in the processing of the
data are: flux calibration and opacity correction,
deduced from skydip measurements and observations of primary and
secondary calibrators; flagging of bad bolometers; 
correlated noise removal; despiking; low-frequency filtering.   
As a result of the correction for correlated noise, uniformly 
extended emission (on scales larger than $2\farcm5$), which mimics 
the variations of the sky emission (skynoise), is filtered out.

The calibration uncertainty in the final map is of the order of 15\%. 
The pointing uncertainty is $\sim4\arcsec$ (0.025~pc at the distance of 
RCW~120). We estimate the rms noise of the final image to be 0.02~Jy/beam. 

\subsection{Spitzer-MIPS observations at 24~$\mu$m and 70~$\mu$m}

\begin{figure*}[htp]
\begin{center}
  \includegraphics[angle=0,width=140mm ]{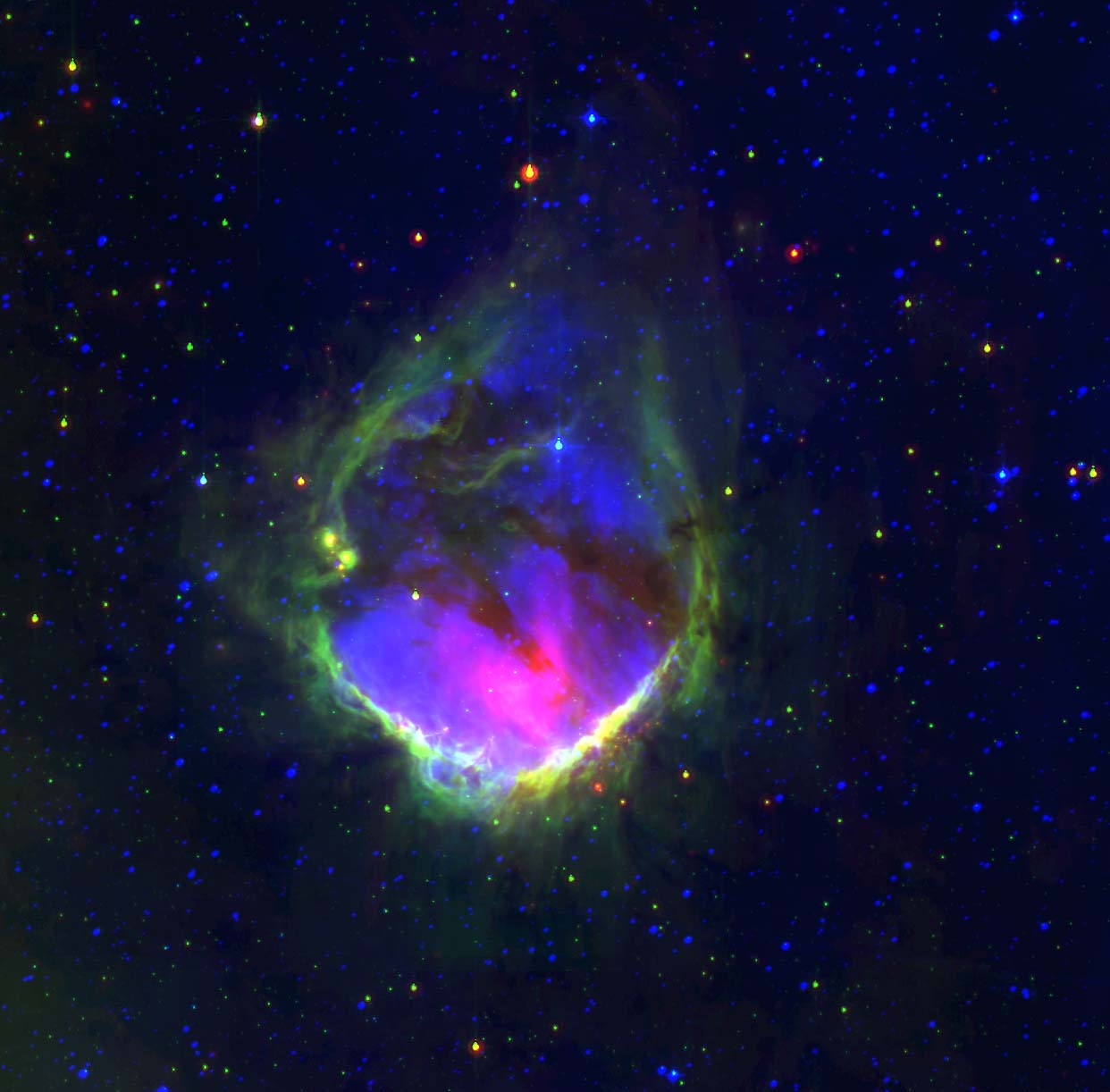}
  \caption{Colour composite image of RCW~120. Blue is H$\alpha$ 
  emission, green is the PAH 
  emission at 8.0~$\mu$m and red is the 24~$\mu$m emission of small 
  dust grains (see text). North is up and east is left. The field size is 
  $\sim$ $24\arcmin$ (N--S) $\times$ $25\arcmin$ (E--W).}
  \end{center}
  \label{midir}
\end{figure*}

Fig.~\ref{midir} is a composite colour image of RCW~120, from the optical 
to the mid-IR. Blue is the H$\alpha$ emission of the ionized gas, from 
SuperCOSMOS (Parker et al.~\cite{par05}). Green is the 8.0~$\mu$m emission 
from Spitzer-GLIMPSE (Benjamin et al.~\cite{ben03}). 
At this wavelength we mainly see the emission of polycyclic 
aromatic hydrocarbons (PAHs). These big molecules are destroyed inside 
the \HII\ region by the UV radiation of the exciting stars. They are excited 
in the photodissociation region (PDR) by the radiation leaking from the 
\HII\ region. Thus they are good tracers of ionization fronts. Red 
shows the emission at 24~$\mu$m, from Spitzer-MIPSGAL 
(Carey et al.~\cite{car05}). We see both the extended emission of the 
small dust grains present inside the ionized region and the PDR, and the 
emission of grains present in the envelopes of YSOs and evolved stars, 
hence the usefulness of the 24~$\mu$m images for detecting YSOs. (The central 
24~$\mu$m extended emission is saturated.)

We have measured three Post BCD frames from the Spitzer-MIPSGAL survey at
24~$\mu$m. These cover a zone going from
$257\fdg85$ to $258\fdg35$ in right ascension and 
from $-38\fdg70$ to $-38\fdg20$ in declination,
encompassing the zone of influence of RCW~120. We used the
DAOPHOT stellar photometry package with PSF fitting (Stetson
\cite{ste87}). The main difficulty to overcome is the measurement of faint
stars superimposed on a bright and very irregular background. For
this we used an iterative process. A first run of DAOPHOT, using the
normal procedures, yields stellar positions and rough magnitudes.
These stars are then subtracted from the original frame. The
resulting image is smoothed with a median filter (window of
11$\times$11 pixels, larger than the size of a star) to remove any
vestiges of the stars, and obtain an image of the background
emission alone. This background image is then subtracted from the
original frame. On the resulting image the stars lie on a low 
brightness and more
uniform background (the only remaining structures are those smaller
than the filter window). Further reduction is performed by the
normal DAOPHOT procedures. This iterative process can be repeated
several times until convergence is satisfactory. We check
the result by eye; if the reduction is good no brightness peaks or holes
should be seen at the positions of the stars after these stars have
been subtracted from the original frame.

The 24~$\mu$m frames overlap, providing us with
double values of the magnitudes of 27 sources. This allows us 
to estimate the accuracy of the results: the rms magnitude difference 
for these 27 pairs is 0.11~mag. 
The DAOPHOT PSF magnitudes are then converted into aperture
magnitudes, following the ``Quick Point Source Photometric Measurement''
procedure (http://ssc.spitzer.caltech.edu/archanaly/quickphot.html), 
and  using isolated bright stars. The resulting
24~$\mu$m magnitudes, obtained for 138 sources, are
given in Table~\ref{photometry}. 

We have also reduced three 70~$\mu$m Post BCD frames, from the 
Spitzer-MIPSGAL survey. Six sources have been measured manually, 
using aperture photometry. These measurements are also given 
in Table~\ref{photometry}.

We have used the interactive software sky atlas Aladin\footnote{
http://aladin.u-strasbg.fr/} to superimpose the 2MASS catalogue 
on the GLIMPSE and 24~$\mu$m images. In the few cases where several 
GLIMPSE or 2MASS sources were observed in the direction of a 
24~$\mu$m source, only the central one was considered. 

Table~\ref{photometry}, which is available in 
electronic form at the CDS, 
presents our measured 24~$\mu$m magnitudes along with
these objects' 1.25~$\mu$m to 8.0~$\mu$m magnitudes taken
from the 2MASS and Spitzer-GLIMPSE catalogues. 
Column 1 gives an identification number. Columns 2 and 3 are the
coordinates according to the Spitzer-GLIMPSE catalogues. Columns 4 to
10 give the $J$, $H$, $K$, [3.6], [4.5], [5.8], and [8.0]
magnitudes. These are extracted from the (highly reliable) 
GLIMPSE~I Spring~'07
Catalog. When necessary they have been supplemented
with values from the (less reliable) GLIMPSE~I Spring~'07 Archive,  
and by the 2MASS All-Sky Point Source Catalog
(a colon indicates an uncertain measurement). Columns 11 and 12 
give our 24~$\mu$m and
70~$\mu$m measurements. Five bright sources which have $[24] \leq 1.38$~mag
(as indicated by a colon) are clearly saturated at 24~$\mu$m.
Column 13 gives our conclusions concerning the evolutionary
stages of the sources (see Sect.~4).

\begin{table}[htp]
\caption{Photometry of the 24~$\mu$m sources. Table~1 is available 
in electronic form at the CDS via anonymous ftp to 
cdsarc.u-strasbg.fr (130.79.128.5) or via 
http://cdsweb.u-strasbg.fr/cgi-bin/qcat?J/A+A/}
\label{photometry}
\end{table}

\section{The distribution of neutral material around RCW~120}

\begin{figure*}[htp]
  \includegraphics[angle=0,width=180mm ]{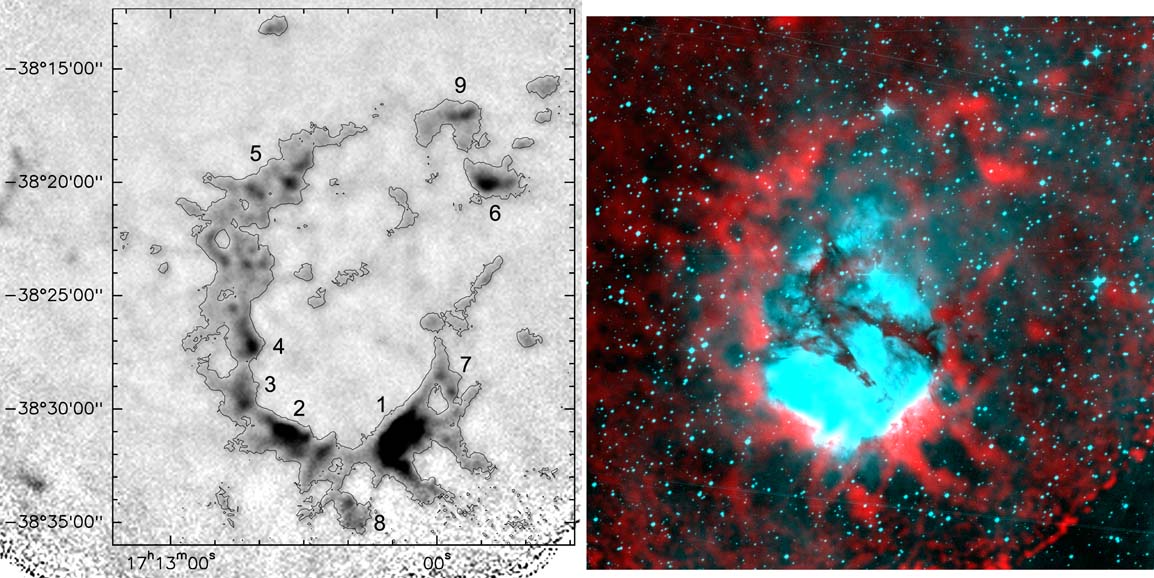}
  \caption{{\bf Left:} 870~$\mu$m emission, observed with APEX-LABOCA. 
  The contour corresponds to the 5\,$\sigma$ level of 0.1~Jy/beam. 
  The dust condensations are identified by a number used in ZA07 
  and in Table~\ref{massemm}. {\bf Right:} Colour composite image of 
  RCW~120: the 870~$\mu$m emission appears in red, the H$\alpha$ emission 
  in turquoise.} 
  \label{apexmap}
\end{figure*}

The distribution of the 1.2-mm dust emission was presented and discussed
by ZA07 using SEST-SIMBA data (beam size 24$\arcsec$, 
rms noise $\sim$0.02~Jy/beam, exposure time 10~hr). Here we present,  
in Fig.~\ref{apexmap},    
the distribution of the 870~$\mu$m dust emission, obtained with the 
APEX-LABOCA camera (beam size $19\farcs2$, 
rms noise $\sim$0.02~Jy/beam, exposure 
time $<\,2$~hr).

Emission at 870\,$\mu$m is mainly the thermal continuum of 
cold dust. Contamination from the \hbox{${}^{12}$CO(3-2)} line may
occur, but this emission is expected to be faint (see the discussion in 
Schuller et al. in preparation).

The 870\,$\mu$m map clearly
resembles the 1.2-mm map: a layer of cold dust
surrounds the ionized region; this layer is fragmented, with condensations 
elongated along the ionization front, especially in the south. 
The higher resolution and higher signal-to-noise ratio at 
870~$\mu$m reveal new structures that were only barely
visible in the 1.2-mm map. This is the case for
the radial structures that are observed north-east 
of the ionized region, and for the filaments responsible for the 
absorption in the central parts of RCW~120 {(the central absorbing 
filaments are conspicuous on Fig.~2, right, blue channel). 
Also, small faint condensations
are now revealed, these often being associated with Class~I sources
detected thanks to the GLIMPSE and MIPSGAL surveys (Sect.~4).

\subsection{Mass determination}
Assuming that the 870\,$\mu$m emission is uniquely thermal dust 
emission, we can derive the mass of the
emitting material. According to Hildebrand
(\cite{hil83}) the total (gas+dust) mass of a condensation is related 
to its flux density $S_\nu$ by
\begin{eqnarray*}
M_{\mathrm{(gas+dust)}} =
100\,\,\frac{S_{\mathrm{870\,\mu m}}\,\,
D^2}{\kappa_{\mathrm{870\,\mu m}}\,\,
B_{\mathrm{870\,\mu m}}(T_{\mathrm{dust}})},
\end{eqnarray*}
where $D$ is the distance of the source, $\kappa_{\mathrm{870\,\mu m}}$ is 
the dust opacity per unit mass at 870\,$\mu$m, 
and $B_{\mathrm{870\,\mu m}}(T_{\mathrm{dust}})$
is the Planck function for a temperature $T_{\mathrm{dust}}$. 
We have assumed a gas-to-dust ratio of 100.  The dust temperature is 
taken as 20~K or 30~K, as often assumed for protostellar
condensations (cf.\ Motte et al.\ \cite{mot03}, 
Johnstone et al.\ \cite{joh06}). 
Following Ossenkopf \& Henning (\cite{oss94}) we adopt  
the value $\kappa_{\mathrm{870\,\mu m}}$~=~1.8 cm$^{2}$~g$^{-1}$.
The dust opacity depends strongly on the dust grains' coating (ice
mantles), compactness, coagulation evolution and the ambient density
(see their fig.~5a--c). This value agrees reasonably well with those
adopted by various authors to derive masses from SCUBA 850\,$\mu$m 
emission maps (e.g. Kirk et al.~\cite{kir06}, Young et al.~\cite{you03}). 

Table~\ref{massemm} lists the measured and derived properties
obtained for several structures discussed in the text (Sect.~5). 
Column 1 gives the structure's name. Condensations \#1 
to \#8 are those identified in ZA07 (their fig.~4). Columns 2 and 3 give the
emission peak coordinates and column 4 gives the emission peak level. Columns
5 and 6 give the intensity value used to define the integration 
boundary and hence the flux density. The choice of the  
levels used to delineate the structures and measure their flux densities 
is somewhat arbitrary; they have been chosen in order to separate the 
different structures.  Finally column~7 gives the derived masses  
depending on the adopted temperature, 20 or 30~K; the
lower mass corresponds to the higher dust temperature. The masses are 
highly uncertain. As shown by column~7, the dust temperature uncertainty 
alone results in a mass uncertainty of a factor two. But the dust 
opacity also is uncertain by a factor of two (Henning, Michel, \& 
Stognienko \cite{hen95}), which implies an uncertainty of the 
same order for the mass.

\begin{table*}[htp]
\caption{Mass estimates}
\begin{tabular}{l c c c c c c}

\hline\hline
 Number & \multicolumn{2}{c}{Peak position} & $F^{\mathrm{peak}}_{\mathrm{870\,\mu m}}$ & Boundary level & $F^{\mathrm{int}}_{\mathrm{870\,\mu m}}$  & Mass range$^{\footnotesize 1}$ \\
        &            $\alpha_{2000}$ & $\delta_{2000}$   & (Jy/beam)    & (Jy/beam) & (Jy)          & ($M_{\odot}$) \\
  \hline
        &            &      &     &  &   &    \\
Shell          &                                    &                                      &       & 0.1 & 190 & 1920--1100 \\
Condensation 1 & 17$^{\rm h}$ 12$^{\rm m}$ 08$\fs$3 & $-$38$\degr$ 30$\arcmin$ 51$\arcsec$ & 13.33 & 0.2 &  78.7 & 800--460 \\
Cond1 core     & 17$^{\rm h}$ 12$^{\rm m}$ 08$\fs$3 & $-$38$\degr$ 30$\arcmin$ 51$\arcsec$ & 13.33 & 2.5 &  24.5 & 250--143 \\
Cond1 b        & 17$^{\rm h}$ 12$^{\rm m}$ 10$\fs$7 & $-$38$\degr$ 32$\arcmin$ 01$\arcsec$ & 2.59  & 1.2 &   4.5  & 46--26 \\
Cond1 c        & 17$^{\rm h}$ 12$^{\rm m}$ 08$\fs$0 & $-$38$\degr$ 32$\arcmin$ 25$\arcsec$ & 1.35  & 0.7 &   3.0  & 31--18 \\
Condensation 2 & 17$^{\rm h}$ 12$^{\rm m}$ 33$\fs$9 & $-$38$\degr$ 30$\arcmin$ 49$\arcsec$ & 1.92  & 0.2 &  18.8 & 192--110 \\
Cond2 bis     & 17$^{\rm h}$ 12$^{\rm m}$ 25$\fs$9 & $-$38$\degr$ 31$\arcmin$ 50$\arcsec$ & 0.73  & 0.2 &   5.4  & 55.5--32 \\
Condensation 3 & 17$^{\rm h}$ 12$^{\rm m}$ 43$\fs$8 & $-$38$\degr$ 29$\arcmin$ 42$\arcsec$ & 0.75  & 0.2 &   6.6  & 63--38 \\
Condensation 4 & 17$^{\rm h}$ 12$^{\rm m}$ 42$\fs$3 & $-$38$\degr$ 26$\arcmin$ 59$\arcsec$ & 0.99  & 0.2 &   8.6  & 88--50 \\
Cond~4 north   & 17$^{\rm h}$ 12$^{\rm m}$ 46$\fs$0 & $-$38$\degr$ 25$\arcmin$ 25$\arcsec$ & 0.46  & 0.2 &   0.9  & 9--5 \\
Condensation 5 & 17$^{\rm h}$ 12$^{\rm m}$ 33$\fs$0 & $-$38$\degr$ 20$\arcmin$ 00$\arcsec$ & 0.77  & 0.1 &  22.1 & 226--130 \\
Cond5 a        & 17$^{\rm h}$ 12$^{\rm m}$ 33$\fs$6 & $-$38$\degr$ 19$\arcmin$ 55$\arcsec$ & 0.77  & 0.3 &   2.6  & 26.5--15 \\
Cond5 b1+b2$^{\footnotesize 2}$    & 17$^{\rm h}$ 12$^{\rm m}$ 41$\fs$4 & $-$38$\degr$ 20$\arcmin$ 17$\arcsec$ & 0.54  & 0.3 & 1.8  & 19--11 \\
Cond5 c        & 17$^{\rm h}$ 12$^{\rm m}$ 43$\fs$3 & $-$38$\degr$ 23$\arcmin$ 31$\arcsec$ & 0.58  & 0.3 &   1.0  & 11--6 \\
Condensation 6 & 17$^{\rm h}$ 11$^{\rm m}$ 49$\fs$5 & $-$38$\degr$ 19$\arcmin$ 59$\arcsec$ & 1.07  & 0.1 &   8.8  & 90--52 \\
Condensation 7$^{\footnotesize 2}$ &                                    &                                      &       & 0.2 & 2.8  & 28.5--16 \\
Cond7 a1       & 17$^{\rm h}$ 11$^{\rm m}$ 59$\fs$5 & $-$38$\degr$ 28$\arcmin$ 25$\arcsec$ & 0.49  & & \\
Cond7 a2       & 17$^{\rm h}$ 11$^{\rm m}$ 56$\fs$9 & $-$38$\degr$ 29$\arcmin$ 09$\arcsec$ & 0.57  & & \\
Condensation 8 & 17$^{\rm h}$ 12$^{\rm m}$ 20$\fs$0 & $-$38$\degr$ 34$\arcmin$ 05$\arcsec$ & 0.56  & 0.2 &   3.8  & 38--22 \\
Condensation 9 & 17$^{\rm h}$ 11$^{\rm m}$ 54$\fs$2 & $-$38$\degr$ 16$\arcmin$ 53$\arcsec$ & 0.50  & 0.2 &   4.2  & 42--24 \\
  \hline
  \label{massemm}
\end{tabular}\\
\\
{$^{\footnotesize 1}$ Mass calculations performed with $T_{\rm{dust}}$=30 and 20~K for the lower and higher values, respectively}\\
{$^{\footnotesize 2}$ The given flux density corresponds to the sum of the two substructures}\\
\end{table*}

The masses 
we estimate from the 870~$\mu$m emission map are roughly twice those 
obtained by ZA07 from the 1.2-mm map. This discrepancy is partly due to the 
values adopted for the dust opacity, and mostly to the levels used to 
delineate the condensations. The new values are more reliable, as the 
870~$\mu$m map has a much better signal-to-noise ratio; also, 
the structures for which we estimate the masses are better defined in 
this paper.

In order to derive the mass contained in the annular structure that
surrounds the ionized region (referred to as ``Shell'' in
Table~\ref{massemm}) we integrated the flux over the surface
enclosed by the 5\,$\sigma$ level contour (corresponding to 0.1~Jy/beam; 
see Fig.~\ref{apexmap}). The result is a total mass for the shell 
in the range 1200\,--\,2100~$M_{\sun}$. 

Condensation 1 contains a very bright core (emission peak of 13.3~Jy/beam) 
superimposed on a bright elongated plateau at a level $\sim\,2$~Jy/beam. 
This core is resolved, with a deconvolved size of 
$14\farcs6\,\times\,7\farcs7$ (0.095~pc\,$\times$\,0.05~pc; FWHM; 
see Sect.~5.3, condensation 1). 

In addition to the condensations listed in ZA07, a few other structures 
were measured, most of them associated with Class~I sources. They are 
identified in Figs~9 to 14.

\subsection{Column density and extinction}

We have calculated the H$_2$ column density, $N(\mathrm H_2)$, from the surface 
brightness $F_{\mathrm{870\,\mu m}}$, using the formula 
\begin{eqnarray*}
N(\mathrm H_2) =
\frac{100\,\, F_{\mathrm{870\,\mu m}}}
{\kappa_{\mathrm{870\,\mu m}}\,\,
B_{\mathrm{870\,\mu m}}(T_{\mathrm{dust}})\,\,
2.3\,\,m_{\mathrm H}\,\,\Omega_\mathrm{beam}}\mathrm,
\end{eqnarray*}
where $F_{\mathrm{870\,\mu m}}$ is expressed in 
Jy~beam$^{-1}$, $B_{\mathrm{870\,\mu m}}$ in Jy, 
$N(\mathrm H_2)$ is per square centimetre, the hydrogen atom mass 
$m_{\mathrm H}$ is in grams and the beam solid angle 
$\Omega_\mathrm{beam}$ is in steradians. 

Adopting $\Omega_\mathrm{beam} = 9.817~10^{-9}$~sr  
corresponding to a beam of $19\farcs2$ (FWHM), and assuming 
$T_{\mathrm{dust}} = 20$~K, this gives
\begin{eqnarray*}
N(\mathrm H_2) = 3.136~10^{22}\,\, F_{\mathrm{870\,\mu m}}.
\end{eqnarray*}

From the classical relations 
$N(\mathrm H+H_2)/E(B-V)=5.8~10^{21}$~ particles~cm$^{-2}$~mag$^{-1}$ 
(Bohlin et al.~\cite{boh78}) and $A_V  = 3.1~E(B-V)$, we obtain 
$A_V  = 5.34~10^{-22}~N(\mathrm H_2)$. 

The bright core in condensation~1, with its emission peak of 13.3~Jy/beam,
is responsible for a visual extinction of 225~mag (assuming a dust 
temperature of 20~K). Similarly, all the material enclosed by the 
5\,$\sigma$ contour (Fig.~\ref{apexmap}, 0.1~Jy/beam level)  
corresponds to a visual  extinction greater than 1.7~mag. Dust filaments~--
very conspicuous in front of the ionized region~--
are responsible for 
the absorption observed at optical wavelengths 
in the direction of the centre of RCW~120 (e.g.\ at H$\alpha$, in 
Fig.~\ref{apexmap}). The material observed at 
870~$\mu$m in the direction of the exciting star of the \HII\ region has a 
small column density, corresponding to a visual extinction 
$\leq\,1$~mag. This is less than the extinction,  
$A_V=4.65$~mag, determined by Avedisova \& Kondratenko~(\cite{ave84}) for 
the exciting star.  
The relatively large beam of APEX-LABOCA smoothes out small-scale 
local features of the dust emission. Also, large-scale structures 
are removed during the reduction (Sect.~2). The visual extinction is 
probably larger than estimated using APEX data, by some 3\,--\,4~mag.

\begin{figure}[htp]
\includegraphics[width=90mm,angle=0]{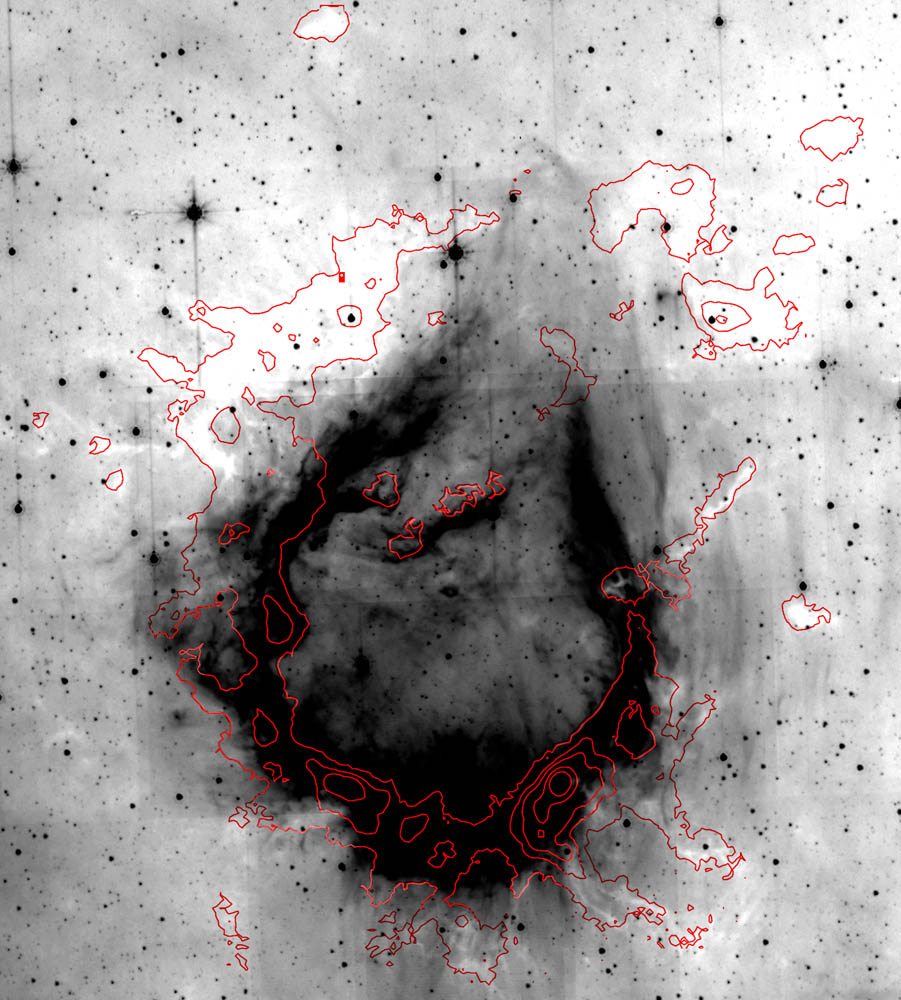}
\caption{Dust condensations observed at 870~$\mu$m and IR dark clouds 
in the vicinity 
of RCW~120. The red contours (0.1, 0.2, 1.0, 2.5, and 5.0~Jy/beam) correspond 
to the 870~$\mu$m emission. The grey background is the Spitzer-GLIMPSE 
image at 8.0~$\mu$m, showing several IRDCs. The most massive 
condensations (in the south) are not IRDCs.}
\label{IRDC}
\end{figure}

The Spitzer emission maps at 8.0~$\mu$m and at 24~$\mu$m show, in absorption, 
several  
infrared dark clouds (IRDCs) in the vicinity of RCW~120. When we compare these 
maps with the dust emission map at 870~$\mu$m, Fig.~\ref{IRDC}, 
we see that most of the 
cold dust condensations in the north have IRDCs as counterparts, whereas 
in the south the brighter cold dust condensations have no IRDC counterparts. 
This gives an indication of the respective positions of the dust 
condensations and the ionized gas: in the north, the dust 
condensations are in front of the PDR, whereas they are  
slightly behind the ionization front in the south. 

\begin{figure*}[htp]
\includegraphics[width=180mm,angle=0]{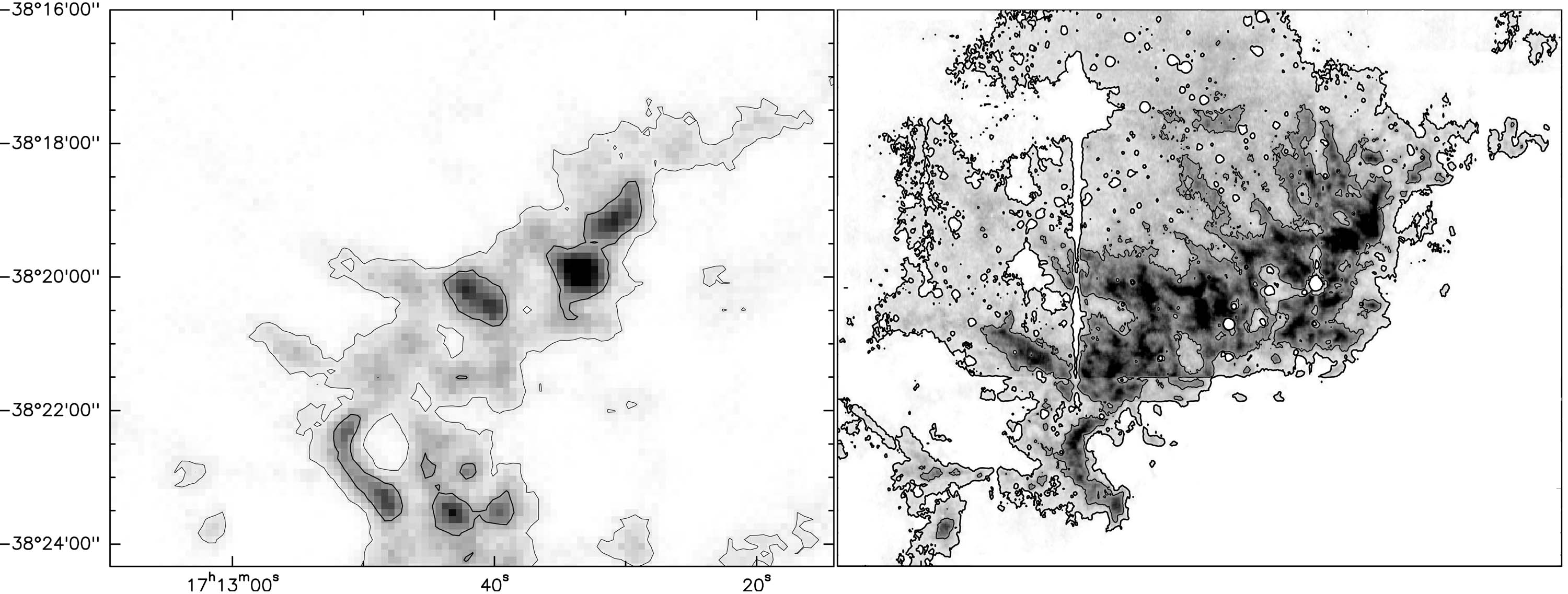}
\caption{Visual extinction in the direction of the 
north-east condensation~5. The grey 
scale is the same for the two figures: white is for 
zero local extinction, black is for an extinction of 10~mag or larger.
{\bf Left:} extinction estimated from the 870~$\mu$m emission. {\bf Right:} 
extinction estimated from the 8.0~$\mu$m absorption. In each figure the 
contours correspond to an extinction of 1.7~mag and 5~mag.}
\label{extinctionNE}
\end{figure*}

The visual extinction can be estimated from the 8.0~$\mu$m absorption 
(cf.\ Vig et al.~\cite{vig07}), and 
compared with that found from the 870~$\mu$m emission. The  
Indebetouw et al.~(\cite{ind05}) extinction law gives 
$A_{8.0\,{\mu}m}/A_K=0.43$; the Rieke \& Lebofsky~(\cite{rie85}) 
extinction law gives $A_K/A_V =0.112$. Thus 
$A_V =22.55~\tau_{8\,{\mu}m}$, with $\tau_{8\,{\mu}m}$ the 
optical depth at 8.0$\,\mu$m. If $F$ and $F_0$ are the 8.0~$\mu$m 
surface brightnesses, respectively in the direction of the absorbing cloud 
and in the direction of the surrounding background (assumed uniform), then 
$\tau_{8\,{\mu}m}=\ln(F_0/F)$. Of course, this method for deriving the 
extinction does not work if the absorbing structure has too great an 
optical thickness; an $A_V$ of 50~mag seems a reasonable maximum value.

Fig.~\ref{extinctionNE} allows us to compare these two 
determinations of the extinction in the direction of a zone 
situated north-east of RCW~120 (in the direction of condensation~5). 
For this region we have 
estimated $F_0$ to be equal to 45~MJy/sr. The first contour, on each map, 
corresponds to a visual extinction of 1.7~mag. These two maps 
are similar, with a few differences:
\begin{itemize}
 \item The region with $A_V \geq 1.7$~mag 
 is more extended on the 8.0~$\mu$m extinction map than on the 
 870~$\mu$m map. In the low extinction zones the extinction 
 derived from the dust emission at 870~$\mu$m
 is 1 -- 2~mag less than that derived from the 8.0~$\mu$m absorption. 
 In the large extinction zones the two determinations of the 
 extinction are in rather good agreement.
 \item The 8.0~$\mu$m extinction map shows small-scale structures,
 mainly filaments and curls. In the 870~$\mu$m observations these
 structures have been smoothed out because of the larger 
 beam. Note that many of the
 large extinction filaments are perpendicular to the ionization front; 
 we will return to this point in Sect.~5.4.
\end{itemize}

An IRDC is also observed in the 24~$\mu$m image, in the same direction.
Because the surrounding background 
is far from uniform, we did not try to estimate 
the extinction via the absorption at 24~$\mu$m. 

What can we learn about the extinction from the near-IR data 
(from the 2MASS survey)? This data is difficult to 
interpret in terms of extinction. Zones devoid of stars, hence 
directions of large extinction, are clearly present all around 
RCW~120 (ZA07, fig.~3),  in the $J$, $H$ and $K$ maps.  
This is the case, for example, of the north-east large  
extinction zone shown in Fig.~3. Very few stars are 
detected towards these large-extinction filaments, 
whereas elsewhere numerous stars are observed, many of them 
displaying large extinction (up to 30~mag in $A_V$) in 
the $J-H$ versus $H-K$ diagram
(not shown here). These are most probably evolved stars situated in the distant 
background, possibly in the Galactic bulge (RCW~120 is less than $12^\circ$  
away from the Galactic centre).


\section {Young stellar objects in the vicinity of RCW~120}

We wish to determine whether each source detected at 24~$\mu$m is a  
star -- main-sequence or evolved -- or is a YSO -- Class~I if its 
luminosity is dominated by an accreting envelope or Class~II if its 
luminosity is dominated by a disk. (Robitaille et al.~\cite{rob06}  
prefer to describe YSOs as stage~I or stage~II sources.)

In the absence of a spectroscopic signature, 
several indicators can be used to establish the evolutionary status
of a YSO, depending on the wavelength range.
\begin{itemize}
\item One indicator is the presence of a near-IR colour
excess in the $J-H$ versus $H-K$ diagram. This excess of near-IR
emission is generally attributed to hot dust associated with a disk
(Lada \& Adams \cite{lad92}).
\item Another indicator often used with the Spitzer-GLIMPSE survey 
is the position of the source in the
Spitzer-IRAC colour-colour diagram, [3.6]$-$[4.5] versus [5.8]$-$[8.0]. 
In this diagram, unreddened stars (both main-sequence and evolved)
have colours near (0,0), whereas Class~I and Class~II sources occupy 
different regions, as shown by Allen et al.~(\cite{all04}) and discussed by 
Whitney et al.~(\cite{whi03a}, \cite{whi03b}, \cite{whi04}). 
\item The slope of the spectral energy distribution (SED) 
between the near-IR and the mid-IR, for example
from 2~$\mu$m to 10 or 20~$\mu$m, is another indicator. 
This slope is defined as 
$\alpha={\rm d}({\log}(\nu F_{\nu}))/{\rm d}({\log}(\nu))$
In the following we shall consider that Class~I 
sources have an increasing SED, i.e. $\alpha >0$, Class~II sources have 
a decreasing SED with $-1.6 \leq \alpha \leq 0$, and Class~III sources 
have $\alpha <-1.6$. One can use for example the $K$ versus $K-$[24] 
diagram (Rebull et al.~\cite{reb07}\footnote{Rebull et al.~(\cite{reb07})  
also identify ``flat spectrum'' sources, which are intermediate 
between Class~I and Class~II, and correspond to $-0.3< \alpha <+0.3$ 
 ($6.75<K-[24]<8.31$). The domain of flat spectrum sources is 
indicated in Fig.~4}).
Considering sources corrected for the interstellar extinction, 
Class~I objects have $K-[24]>7.54$, and Class~II objects have
$K-[24]$ between 3.37 and 7.54. Class~III and stars, which we do not 
attempt to separate, have $K-[24]<3.37$. (Ordinary stellar photospheres 
have $K-[24]\sim\,0$.)
\end{itemize}

We shall compare the information given 
by these three indicators to discuss the evolutionary stages of the sources
observed at 24~$\mu$m in the vicinity of RCW~120, 
and listed in Table~\ref{photometry}. We shall show the limitations of these 
indicators and try new ones. We have been greatly helped in this study by the 
comprehensive discussion of Robitaille et al.~(\cite{rob06}).

%
\begin{figure}[htp]
\includegraphics[width=90mm,angle=0]{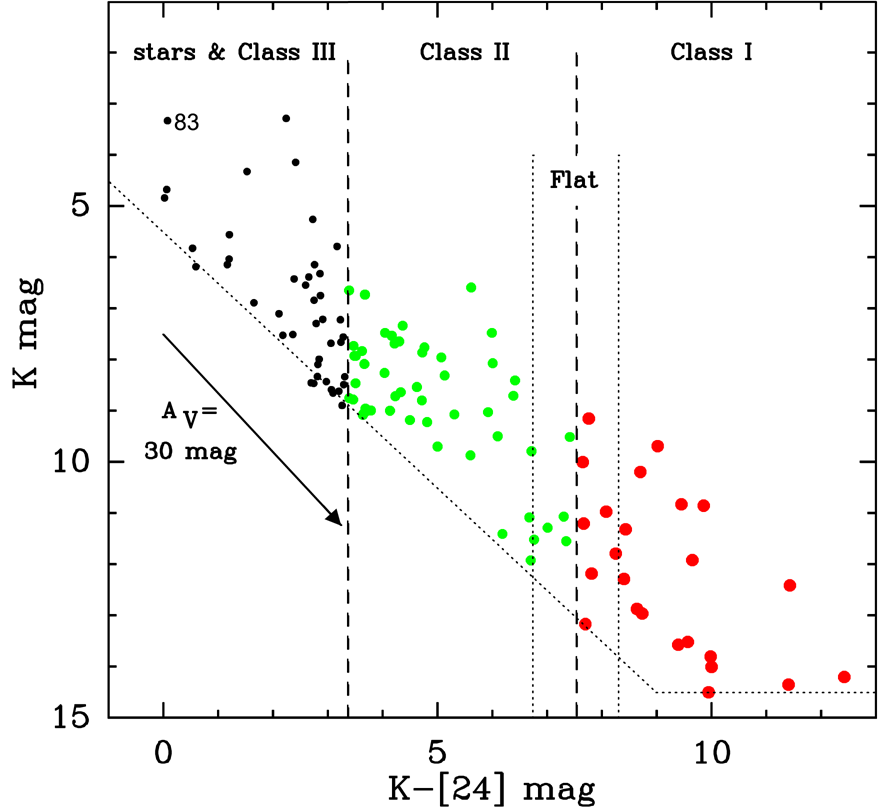}
\caption{The $K$ versus $K-[24]$ diagram. Different classes of sources are
separated by dashed lines. The dotted lines correspond to the detection
limits of 14.5 mag in $K$ and 5.5 mag in [24]. (Note that these
detection limits are not completeness limits.) The black arrow is a reddening line
corresponding to a visual extinction of 30~mag. 
Class~I sources appear as large red dots, Class~II sources 
as medium green dots, and all the other sources as small black dots. 
The same symbols will be used for the same sources in Figs~\ref{cc1}, 
\ref{cc2}, and \ref{hr3}~(left).}
\label{hr1}
\end{figure}

Fig.~\ref{hr1} presents the $K$ versus $K-$[24] diagram obtained for the
118 sources that have $K$ and [24] magnitude measurements.
We are tempted to identify 24 Class~I sources (red symbols),
and 51 Class~II sources (green symbols) in this diagram. However, note
the strong effect of interstellar extinction on the $K-$[24] colour,
and the suspicious concentration of sources near the separation line 
between Class~II and Class~III objects. We shall return to
this point later in this section. (Following Robitaille et al.~\cite{rob06}, we here refer
-- unconventionally -- to the extinction as ``interstellar'' when it is due to
dust located outside the envelope and/or disk of a source.)

Fig.~\ref{cc1} presents the [3.6]$-$[4.5] versus [5.8]$-$[8.0] diagram
for the 103 sources that have been measured in the
four IRAC bands. We use the same symbols as in Fig.~\ref{hr1} 
for the Class~I and Class~II sources already  
identified with the first indicator ($K$ versus $K-[24]$). 
Fig.~\ref{cc1} shows an
almost perfect agreement between the two indicators for Class~I sources. 
Eleven new very red sources (black squares), which have no 
measurable $K$ magnitude and hence are absent in Fig.~\ref{hr1}, 
are identified as Class~I. 
The situation is less satisfactory for the candidate Class~II sources  
(according to our first indicator). A number of these 
sources lie inside the ellipse of stars. We shall return to this 
point later in this section.


\begin{figure}[htp]
\includegraphics[width=90mm,angle=0]{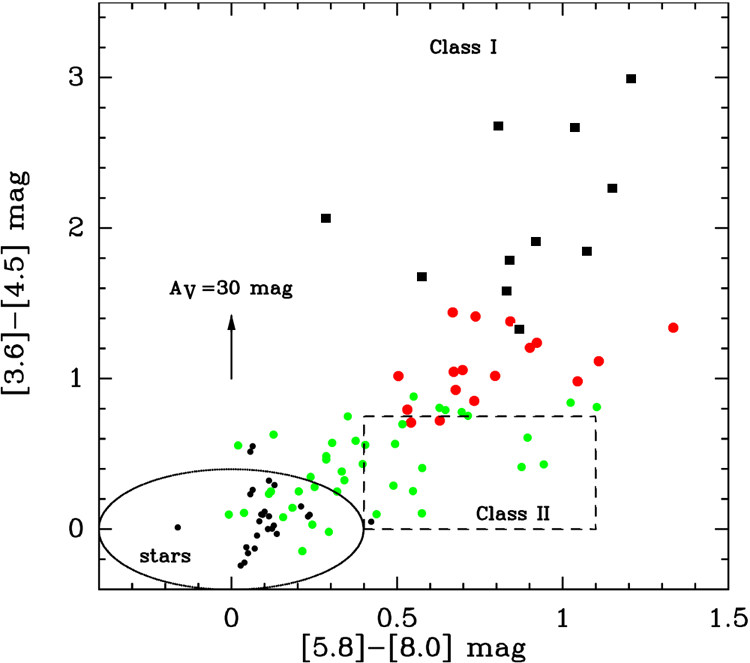}
\caption{The $[3.6]-[4.5]$ versus $[5.8]-[8.0]$ diagram. 
The black arrow is a reddening line
corresponding to a visual extinction of 30~mag. Each source is  
represented by the same symbol as in Fig.~\ref{hr1}. The black 
squares are for sources lacking $K$ magnitudes. 
The locations, according to Allen et al.~(\cite{all04}), of sources dominated
by their photospheres (stars and Class~III objects) and of Class~II sources are
indicated by an ellipse and a rectangle, respectively.}
\label{cc1}
\end{figure}
%
\begin{figure}[htp]
\includegraphics[width=90mm,angle=0]{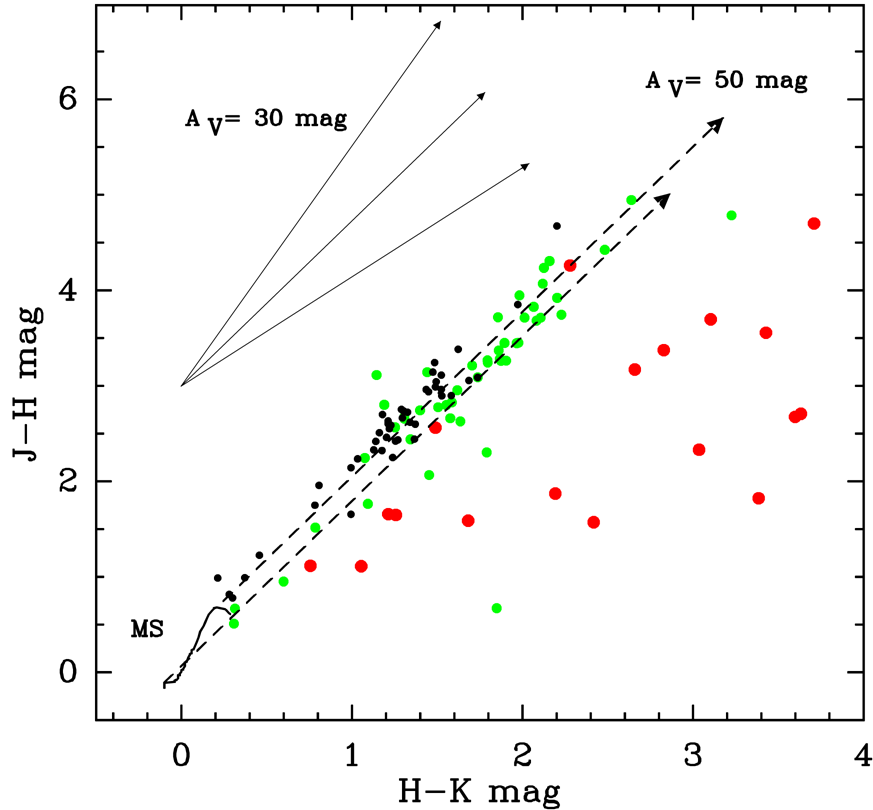}
\caption{The $J-H$ versus $H-K$ diagram. The main
sequence (MS) is drawn with a solid curve;  the
colours are from Martins \& Plez (\cite{mar06}) for O stars,
and from Tokunaga (\cite{tok00}) for later spectral types. The reddening
lines (dashed lines), for O5V and M2V stars,
are for a visual extinction of 50~mag (extinction law of
Indebetouw et al.~\cite{ind05}). The symbols are the same as
in Fig.~\ref{hr1}. All the sources situated below the lower (O-star)
reddening line exhibit a near-IR excess. The sources situated
above the upper reddening line are possibly evolved stars. The 
solid-line arrows are the reddening lines corresponding to a 
visual extinction of 30~mag and compatible with the extinction law of 
Indebetouw et al.~(\cite{ind05}; see text).}
\label{cc2}
\end{figure}

Fig.~\ref{cc2} is the $J-H$ versus $H-K$ diagram for the 112 sources
with measured $J$, $H$, and $K$ magnitudes. All the 
Class~I sources (identified by our first indicator) exhibit  
near-IR excesses, and they are nearly the only ones to do so. (The relative 
amounts of the near-IR excesses is uncertain, as almost all these 
sources have uncertain 2MASS magnitudes.) The fact
that most of the sources previously classified as Class~II do not
have near-IR excesses is surprising, since in models of Lada \&
Adams (\cite{lad92}), the hot dust situated in the disk
emits in the near-IR. One possible explanation is that the slope of the
extinction curve is steeper than that adopted in Fig.~\ref{cc2}.
For this figure we have used the interstellar extinction law of
Indebetouw et al.~(\cite{ind05}). According to these authors,
$A_J/A_K=2.50\pm0.15$ and $A_H/A_K=1.55\pm0.08$, giving a slope
$(A_J-A_H)/(A_H-A_K)=1.73\pm0.67$. This range of values is illustrated 
in Fig.~\ref{cc2}, showing that a small increase of the slope would result
in a near-IR excess for numerous Class~II sources.

The sources with $K-[24]<3.37$ are Class~III objects or
stars -- in the later case mainly evolved stars. Main-sequence 
stars at the distance of
RCW~120 or in the background (thus possibly reddened) should have
$K>5.67$~mag (Martins \& Plez \cite{mar06}). Thus all
the sources brighter than 5.67 in $K$ and with $K-[24]<3.37$ are
possibly giants or supergiants. This is compatible with their positions in
the $J-H$ versus $H-K$ diagram. One of these sources, \#83, is 
very bright in $K$ and lies in the direction of the ionized gas. 
This star is HD\,155275, an M0III star (Houk \cite{hou82}; it is identified 
in Deharveng \& Zavagno \cite{deh08}).

Many sources may be affected by a large interstellar 
extinction. Fig.~\ref{cc2}, for example, shows that many stars 
(black dots) or candidate Class~II sources (green dots) may be affected 
by an (external) extinction between 20 and 40~mag. This extinction 
may severely affect the classification of the sources. We suggest 
that some stars or Class~III objects, affected by strong 
extinction, are mistaken for Class~II objects by our first indicator. 
Our first indicator shows a large concentration of stars and Class~III 
and Class~II 
objects at $K-[24]\sim3.37$. All the Class~II sources 
which have $3.37<K-[24]<4.0$ lie inside the ellipse (showing the
location of stars)
in Fig.~\ref{cc1}. Furthermore, according to their $J-H$ and $H-K$ colours  
(Fig.~\ref{cc2}), all these sources may be evolved
stars. Thus most of these stars are probably evolved background stars 
and not Class~II sources.

And the situation is perhaps not better for Class~I sources.
Fig.~\ref{hr3} presents the [24] versus [8]$-$[24] 
diagram, very similar to the $K$ versus $K-$[24] diagram of Fig.~\ref{hr1}, 
with the disadvantage of a smaller wavelength range but the advantage of a 
weaker influence of the extinction. In this diagram unreddened 
Class~I sources have $[8]-[24]>3.58$, and Class~II sources have 
$1.67 \leq [8]-[24] \leq 3.58$. We have kept the same 
symbols as given by our first indicator.  Fig.~\ref{hr3} (left) 
shows that some 
sources previously classified as Class~I now appear as Class~II, 
and that some sources previously classified as Class~II are probably 
Class~III sources or stars.

%
\begin{figure*}[htp]
\includegraphics[width=180mm,angle=0]{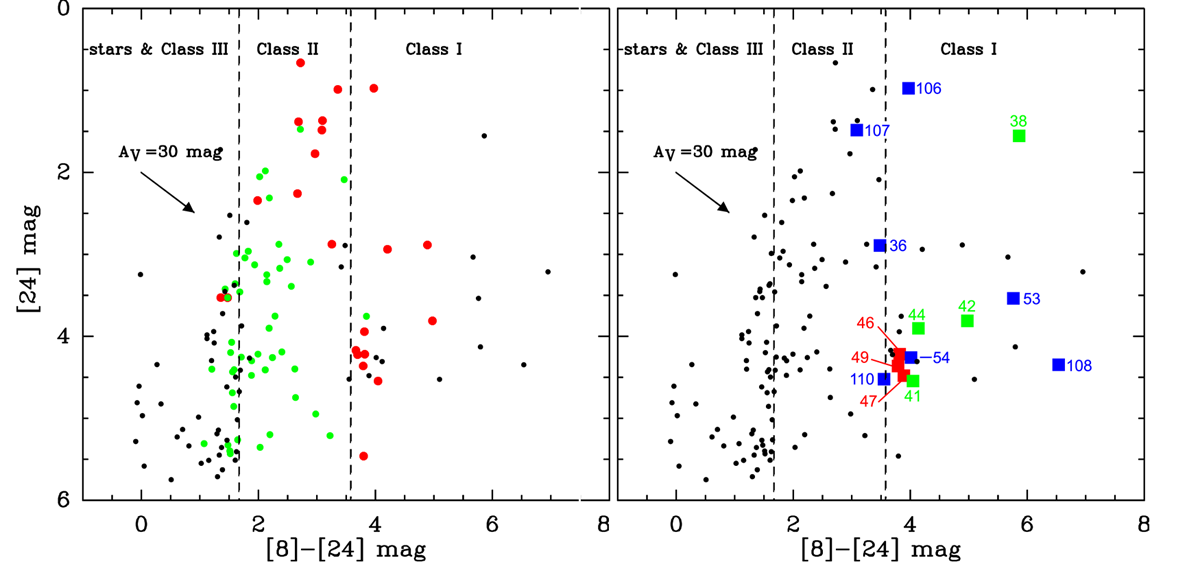}
\caption{The [24] versus [8]$-$[24] diagram, with the different classes 
separated by dashed lines. Class~I sources have  
[8]$-$[24]$ \geq 3.58$, Class~II sources have $1.67\leq$ [8]$-$[24] $ \leq 3.58$. 
The reddening vectors are for a visual extinction of 30~mag.
{\bf Left:} The red and green dots correspond to Class~I and 
Class~II sources, 
according to our first indicator. {\bf Right:} The maximum external
extinction possibly affecting the sources is indicated as follows:  
red squares are for sources with $A_V(\mathrm {external}) \geq 30$~mag, green 
squares for sources with 20~mag $\leq A_V(\mathrm {external})\leq 30$~mag, blue  
squares for sources with 5~mag $\leq A_V(\mathrm {external}) \leq 10$~mag 
(no sources with 10~mag $\leq A_V(\mathrm {external}) \leq 20$~mag are present 
in this diagram).}
\label{hr3}
\end{figure*}

We shall now try to estimate the possible influence of the 
interstellar extinction on our classifications. We use 
the 870~$\mu$m emission to derive a maximum value of this external 
extinction; we assume that the sources lie in the centres of the 
dust condensations observed at 870~$\mu$m in their direction. The 
extinction is then given by 
\begin{eqnarray*}
A_V(\mathrm {external}) = \frac{16.75\,\, F_{\mathrm{870\,\mu m}}}{2}+3,
\end{eqnarray*}
where $F_{\mathrm{870\,\mu m}}$ is expressed in Jy/beam  
and the 3 is to correct for a 3~mag underestimate of the 
extinction using the 870~$\mu$m map (Sect.~3.2). Fig.~\ref{hr3} 
(right) gives an idea of the maximum external extinction possibly 
affecting the sources. It shows that only a few sources are possibly  
misclassified for this reason. These are mainly sources observed 
in the direction of the bright condensation~1 (\#41, \#44, \#46, 
\#47, \#49, \#54) and in condensation~6 (\#106, \#110) (Sect.~5.3). 
These sources 
may be flat spectrum sources. For the other sources, the discrepancy 
between the classifications given by different indicators is most 
probably due to difficult photometric measurements. It mainly affects 
sources lying near the separation limits of the different classes.

Table~\ref{photometry} gives our conclusions concerning the evolutionary 
stages of the sources. As Class~I we include 21 sources
satisfying all the indicators ($K$ versus $K-[24]$, 
$[3.6]-[4.5]$ versus $[5.8]-[8.0]$, [24] versus [8]$-$[24]) and not 
affected by too much external extinction. We consider as intermediate 
Class~I--Class~II 18 sources which are Class~I or Class~II 
depending on the indicator (identified by ``I-II?'' in column 13 of Table~1), 
or are possibly flat spectrum sources mistaken for 
Class~I because of a large external extinction (``I-flat?'' in Table~1)). 
We consider as Class~II 
29 sources satisfying all the indicators, and as intermediate 
Class~II--Class~III 9 sources which are Class~II or Class~III 
depending on the indicators (``II-III?'' in Table1).
Note that in Table~\ref{photometry} we have classified as ``stars'' 
all the sources dominated by their photosphere, and which we 
are unable to distinguish: Class~III YSOs, MS or evolved stars.

The spatial distribution of the YSOs is presented on Fig.~\ref{classification}  
and is discussed in Sect.~5.3.

\section{Discussion}
\subsection{Dust condensations and infrared dark clouds}

RCW~120 demonstrates in an exemplary way that massive dust 
condensations are not all infrared dark clouds. A dense 
neutral condensation needs to be seen 
towards a bright background emission to appear as an IRDC. But 
another necessary condition is to not be located behind bright 
foreground emission. In the case of RCW~120 the structures that 
are the most massive (condensations 1 and 2) and the densest 
(the core in condensation 1, responsible for an extinction $\sim 225$~mag) 
are not IRDCs: they are masked by the bright emission of the 
adjacent foreground PDR. Only their external parts 
(those far from the IF) appear as faint absorption features  
(Fig.~\ref{IRDC}).

\subsection{Young stellar objects and AGB stars} 

Robitaille et al.~(\cite{rob08}) present a catalogue of intrinsically 
red sources, observed by Spitzer, in the Galactic mid-plane. These 
sources consist mostly of high- and intermediate-mass YSOs and 
asymptotic giant branch (AGB) stars. As shown by these authors, 
it is very difficult to distinguish between these two types.

In the following we compare the sources detected at 24~$\mu$m in the 
vicinity of RCW~120 with  the red sources selected by Robitaille et al. 
If we apply their selection criteria, 
i.e. [4.5]$-$[8.0]$\geq 1$, $13.89 \geq [4.5] \geq 6.5$, and 
$9.52 \geq [8.0] \geq 4.01$, to the 112 sources with [4.5] and [8.0] 
magnitudes in Table~1, we find 39 sources; 
26 of them are Class~I or flat spectrum sources, 11 are 
Class~II, and two more are uncertain Class~II--Class~III. 

Robitaille et al.\ estimate that $\sim$19 percent of the red sources 
may be AGBs. If true, this means that up to seven of 
our 39 red sources may be AGBs and not YSOs! It is very 
difficult to distinguish between these two categories if we only have
photometric measurements up to 24~$\mu$m. Robitaille et al.\ show that 
YSOs are generally redder than AGBs between 8~$\mu$m and 24~$\mu$m, and 
propose to use the criterion [8.0]-[24]$\geq 2.5$ to identify YSOs. 
Using this criterion we find that 32 of our 39 red sources are 
probably YSOs (the seven sources with [8.0]-[24]$\leq 2.5$ are
Class~II or uncertain Class~II). Thus we must keep in mind in the 
following that a few of our sources identified as YSOs, and especially 
Class~II YSOs, are possibly AGB stars and are therefore evolved.

\subsection{Large scale distribution of the YSOs}

We have measured 138 sources around RCW~120 emitting at 24~$\mu$m.  
Three are unclassified. Twenty-one are Class~I YSOs, and eighteen are 
possible Class~I or flat spectrum sources. Twenty-nine sources are 
Class~II YSOs, and nine   
sources are more uncertain Class~II sources. Thus 49 to 56 percent 
of the 24~$\mu$m sources are young stellar objects, 28 percent 
of the 24~$\mu$m sources being Class~I or flat spectrum sources.  

Fig.~\ref{classification} shows the large-scale distribution of the YSOs 
possibly associated with RCW~120. It shows a large number of Class~I 
or flat spectrum sources (red symbols), all but five associated with 
the shell of dense neutral material surrounding RCW~120. Thus even if 
we have no direct proof that any given YSO is associated with RCW~120, 
this general picture demonstrates that most of them are associated with the 
\HII\ region, and that their formation has been triggered, one way or 
another, by the \HII\ region. On the other hand, the Class~II sources are 
more widely scattered over 
the field. Their association with RCW~120 is less evident, and  
this raises questions about their origin.

We note the large concentration of Class~I or flat spectrum YSOs 
in the direction of condensation~1, the most massive. 

\begin{figure}[htp]
\includegraphics[width=90mm,angle=0]{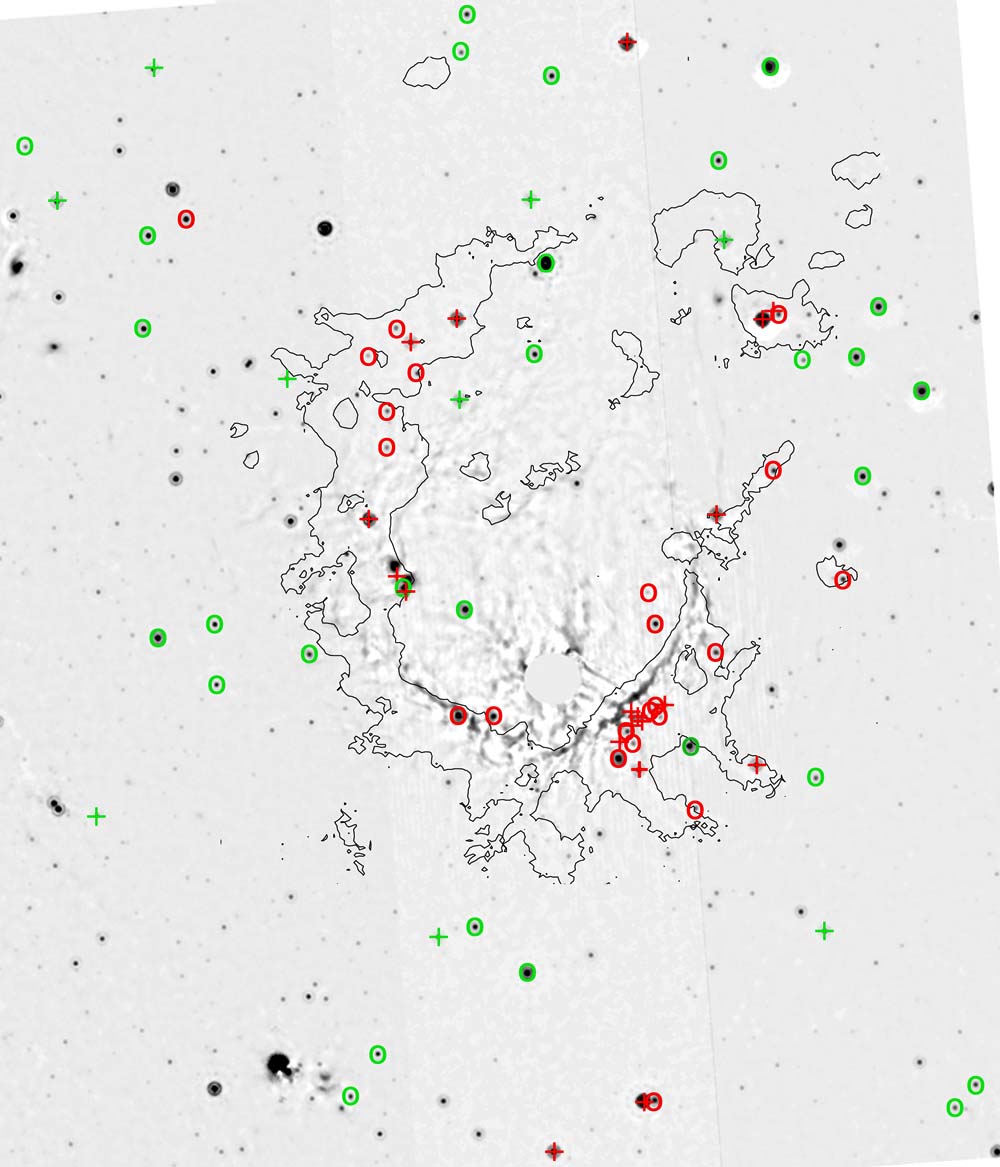}
\caption{Distribution of the Class~I (red circles), intermediate 
Class~I-Class~II or flat spectrum (red crosses), Class~II (green circles), 
and possible Class~II-Class~III sources (green crosses). The grey-scale  
background is an unsharp-masked 24~$\mu$m image, where the extended 
dust emission has been subtracted 
to enhance the stellar component. The contours represent the 870~$\mu$m 
emission level of 0.1~Jy/beam (5\,$\sigma$).}
\label{classification}
\end{figure}

\subsection{Stellar content of the condensations}

In the following we discuss the YSO content of a few condensations.

$\bullet$ {\bf Condensation 1:} This is the most massive of the condensations, 
in the range 460--800~$M_{\sun}$. It contains a very massive core, of  
140--250~$M_{\sun}$, and several other bright substructures. Fig.~\ref{cond1} 
shows this condensation and the numerous 24~$\mu$m sources observed in its 
direction or vicinity. 

The bright 870~$\mu$m core has a size (beam-corrected FWHM) of 
$14\farcs6\,\times\,7\farcs7$ (0.095~pc~$\times$~0.05~pc), along the IF 
and perpendicular to it, respectively -- thus slightly elongated along the IF. 
Its size and peak flux indicate a mean density of a few 
times $10^6$~molecules~cm$^{-3}$. No 
bright 24~$\mu$m source is observed in the direction of the 870~$\mu$m 
emission peak, but three faint sources (including \#47)  
may be associated with it. The Spitzer-MIPSGAL image at 
70~$\mu$m in Fig.~\ref{core1} shows an enhancement of the emission 
in the direction of the core, suggesting that a very deeply embedded 
object is present here. We measured a flux of 130~Jy for this object, 
but this is very uncertain due to the underlying emission of the PDR and 
the low quality of the 70~$\mu$m image. 

According to our indicators, many 24~$\mu$m sources appear to be 
Class~I YSOs. Among these are \#38, which is observed in the direction of a 
spherical dust clump (b in Table~\ref{massemm} and Fig.~\ref{cond1}); \#58, 
also observed on the border of clump a2 in condensation 7; and \#61, 
observed in the direction of the ionized gas 
(but most probably situated behind the \HII\ region). Sources \#36 and \#37, 
also located in projection close to a condensation, are possibly in an 
intermediate stage between Class~I and Class~II. The stage of 
\#39 depends on the indicators used, but is probably a Class~II. 
Source \#32 has only [4.5] and [24] magnitudes; the slope of its SED 
suggests a Class~I source. 

Condensation 1 contains a most remarkable feature: a chain of eleven 
24~$\mu$m sources. They are aligned parallel to the main IF and are regularly 
spaced, separated by some 0.1~pc. They are deeply embedded. Sources 
\#42 and \#53 are Class~I. Sources \#41, \#44, \#46, \#47, \#49, and \#54 
lie near the separation between Class~I and Class~II and may be flat spectrum 
sources as they may be affected by a large external extinction 
(cf. Fig.~\ref{hr3}). Measurements lack for \#40, \#48, and \#52; the slopes of 
their SEDs suggest Class~I sources. 

We have fitted the SEDs of a few sources using the 
Web-based SED fitting-tool of Robitaille et al.~(\cite{rob07}; 
http://caravan.astro.wisc.edu/protostars/). The parameters of the best model, 
as well as a range of possible values for some parameters, are given in 
Table~\ref{models}. YSO \#38 is a stage~I source, with a central object 
of a few solar masses; condensation b is probably not the source's cold 
envelope since its flux density 
at 870~$\mu$m is too large. Source \#61 is also in stage~I, with a central 
object of a few tenths 
of a solar mass. For sources \#53 and \#49, most of the models indicate that 
they are stage~I sources around central objects of about one solar mass, 
and with  massive disks; but a few of the good models (those with 
$\chi^2-\chi_{\rm best}^2 \leq 3$)
show central objects with a higher mass, no accreting envelopes, and 
massive disks, hence corresponding to stage~II sources.

The chain of sources aligned along the IF, and regularly spaced, indicates 
that star formation has occurred in the collected 
layer, probably resulting from gravitational instabilities along a dense filament. 
At a temperature of 20~K and a density of 10$^5$~molecules per cm$^3$, 
gravitational instabilities form cores separated 
by the Jeans length $\lambda_{\rm J} \sim 0.1$~pc, and with the Jeans mass 
$M_{\rm J} \sim 1.5~M_{\sun}$. 
This is probably what has occurred inside condensation~1. On the other hand the 
formation of the very massive and dense core may result from the collect 
and collapse process, i.e. from large-scale 
gravitational instabilities along the surface 
of the collected shell. Here again triggered star formation is probably at work, 
as shown by the presence of a very deeply embedded source -- hence a Class 0 
YSO -- only detected at 70~$\mu$m.

\begin{figure}[htp]
\includegraphics[width=90mm,angle=0]{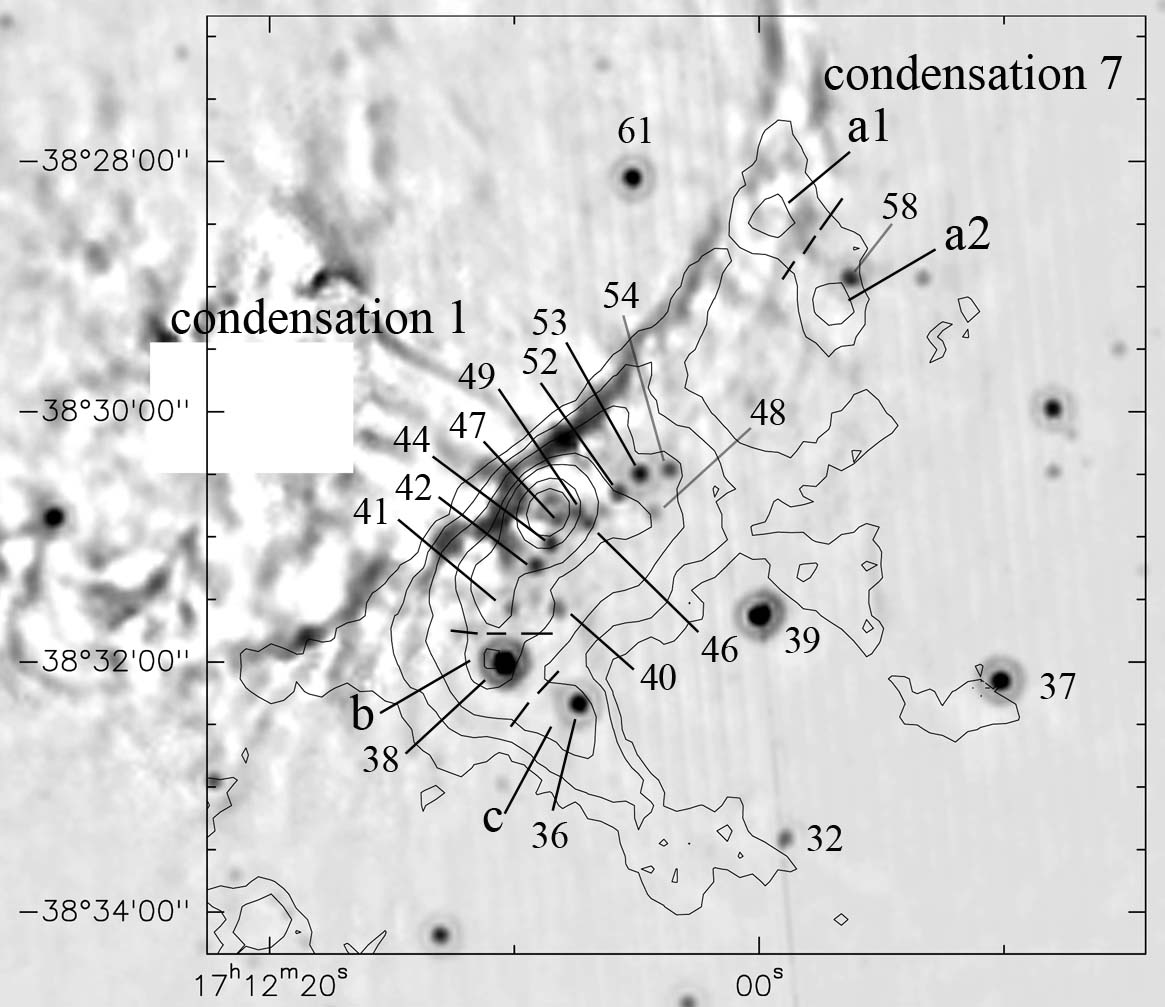}
\caption{Condensations 1 and 7: the background image is an unsharp masked image 
obtained from the Spitzer-MIPSGAL frame at 24~$\mu$m. This image, free of 
the diffuse extended emission, enhances all the stellar sources. 
The contours correspond to the 870~$\mu$m emission. 
The contour levels are 0.2~Jy/beam (used to estimate the masses of 
condensations 1 and 7), 0.4, 0.8, 1.75, 2.5 (used to estimate the mass 
of the core in condensation 1), 5, and 8~Jy/beam.}
\label{cond1}
\end{figure}

\begin{figure}[htp]
\includegraphics[width=90mm,angle=0]{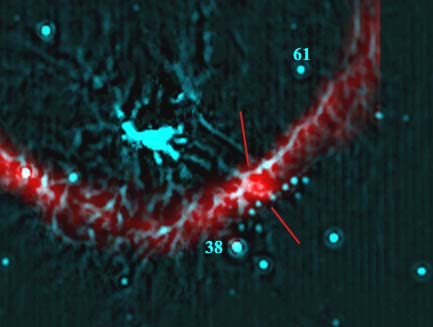}
\caption{Colour composite image of the southern IF in the mid-IR. The 
Spitzer-MIPS emission at 70~$\mu$m appears in red. Turquoise shows an 
unsharp-masked image at 24~$\mu$m, obtained from Spitzer-MIPSGAL. 
The red lines point to the 70~$\mu$m source detected in the exact 
direction of the dense core in condensation~1.}
\label{core1}
\end{figure}

$\bullet$ {\bf Condensation 2:} This lies at the south-east border of RCW~120, 
adjacent to the main IF. Its mass is in the range 110--290~$M_{\sun}$ 
(not including the elongated structure which  we call condensation 2bis 
in Fig.~\ref{cond2}). Its peak column density corresponds to a visual extinction 
$\sim$\,32~mag. The YSO \#50 lies at less than 0.03~pc (in projection) from 
the peak direction.

This source has no near-IR emission (in the 2MASS survey), and 
no measured 8.0~$\mu$m flux, hence does not appear in our diagrams. However 
the slope of its SED between 3.6~$\mu$m and 24~$\mu$m is between 2.0 (without 
extinction correction) and 1.3 (correcting for a visual extinction of 32\,mag). 
Thus source \#50 is a Class~I object. It lies very close 
to the IF, which is probably the reason why we do not detect it at 70~$\mu$m.

Source \#51 lies at the top of a ``finger'', a structure of the IF protruding 
inside the ionized gas, and pointing towards the exciting star. This finger, 
which is seen south-west of condensation 2, is an elongatd dust filament. 
Another  
such structure lies nearby, which we have called condensation 2bis 
(Fig.~\ref{cond2}). ZA07 proposed that these structures result 
from dynamical instabilities of the expanding IF, such as these simulated by 
Garc\'{\i}a-Segura \& Franco (\cite{gar96}). These structures are found towards 
a gap between condensations 1 and 2, through which the ionized gas 
tries to flow away from the central \HII\ region (a region thus dynamically 
unstable). An alternative origin is possible. These two ``fingers'' are 
reminescent of the ``pillars'' in M16, with Class~I \& II YSOs located at 
their tops (Sugitani et al.~\cite{sug02}, Indebetouw et al.~\cite{ind07} 
and references therein). The 
pillars in M16 clearly result from the interaction of 
NGC~6611's OB stars' UV radiation with pre-existing dense clumps. However, 
as shown by the molecular observations of  
White et al.~(\cite{whi99}), most of the mass of 
the pillars is located in their heads rather than their tails, whereas 
the contrary is observed in RCW~120's fingers (at least in condensation~2bis). 
Thus the origin of the fingers and of YSO~\#51 is uncertain; 
higher angular resolution observations are needed.

Our indicators show that YSO \#51 is possibly a Class~I object. 
Its SED, modelled by the Robitaille 
et al.\ fitting tool, confirms this conclusion. The best model indicates a 
central object of 1.3~$M_{\sun}$ with a massive disk but, however, with 
an accreting envelope, hence a stage~I-stage~II object. Several 
parameters are rather uncertain, but the disk is massive in all the models  
(Table~\ref{models}). 

 Finally the other YSO of the field, \#59, is clearly a rather massive 
 stage~II-stage~III source. It is well fitted by a central object of 
 6.6~$M_{\sun}$, a small disk, and no envelope (Table~\ref{models}). 
 Its association with RCW~120 is uncertain. 

\begin{figure}[htp]
\includegraphics[width=90mm,angle=0]{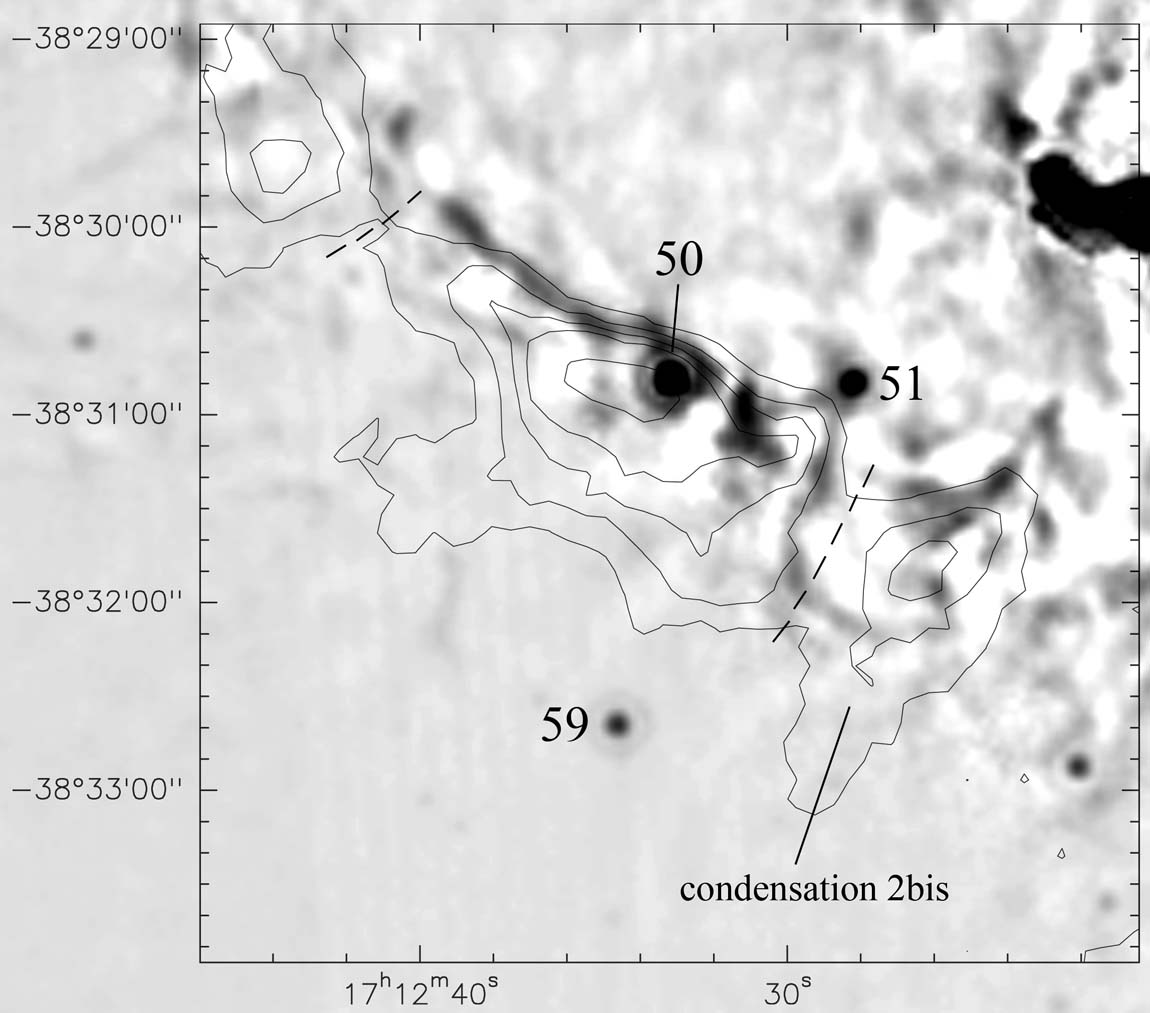}
\caption{Condensation 2: the background image is the unsharp masked image 
at 24~$\mu$m; the contours correspond to the 870~$\mu$m emission. 
The contour levels go from 0.2~Jy/beam (used to estimate the masses of 
the condensations) to 1.8~Jy/beam, by step of 0.2~Jy/beam. The dashed 
lines show the north-east and south-west limits adopted to delimit 
condensation 2. The nearby condensation 2bis is identified.}
\label{cond2}
\end{figure}

$\bullet$ {\bf Condensation 4:} This is elongated along the IF but protrudes inside 
the ionized region. Its mass is between 50 and $90~M_{\sun}$. Fig.~\ref{cond4} 
shows that several 24~$\mu$m sources are detected in its vicinity,
but that none is observed in the 
exact direction of the 870~$\mu$m emission peak. However all three sources 
(\#64, \#67, and \#69) lie on the border of the condensation 
facing the ionized gas and have near-IR counterparts.  
Our indicators show that \#64 is intermediate between Class~I and Class~II, 
and that \#67 is a Class~II. Source \#69 has no measurement at 8.0~$\mu$m, 
so it is not in our diagrams, but the slope of its SED between 3.6~$\mu$m  
and 24~$\mu$m lies between 0.4 (uncorrected for extinction) and 0.1 
(assuming $A_V=14$~mag, corresponding to 
the 870~$\mu$m column density in its direction); thus source \#69 is 
probably intermediate between Class~I and Class~II.

We have used the Robitaille et al.\ SED fitting tool to model source \#67.
The result confirms our previous conclusion. The parameters of the best 
model are given in Table~\ref{models}. This source contains a 
central object of mass $\sim 5~M_{\sun}$. Many of the good models have no accretion from 
an envelope. Thus this source is most probably a stage~II, but 
the mass of the disk is very uncertain.

Two regions of extended emission at 8.0~$\mu$m also lie on the border of 
condensation~4. ZA07 suggested that they were PDRs around Herbig Ae/Be stars. 
Emission from these PDRs is also observed at 24~$\mu$m. Their central stars 
have been observed by 2MASS; their near-IR photometry suggests a B4V star 
affected by a visual extinction of 7.8~mag (object A in Fig.~\ref{cond4}, or 
object 1 in ZA07), and a B7V star with an extinction of 12.2~mag (object B 
in Fig.~\ref{cond4}, or object 2 in ZA07). These stars are probably the 
most massive of the sources associated with condensation~4. 

ZA07 proposed that condensation~4 was a pre-existing dense clump which 
became surrounded by ionized gas during the expansion of the \HII\ region. 
Star formation is observed at its borders and not at its centre. Thus star 
formation is probably not due to the clump implosion (which is predicted 
by the radiation-driven implosion models; Lefloch \& Lazareff~\cite{lef94}, 
Kessel-Deynet \& Burkert~\cite{kes03}).

Fig.~\ref{cond4} shows the presence of a small spherical clump 
north of condensation 4, with a mass in the range 5--9~$M_{\sun}$. A bright 
24~$\mu$m YSO, \#76, coincides with this clump to within the uncertainties 
(about 0.015~pc in projection).  
The evolutionary stage of \#76 is uncertain: our indicators point 
to Class~I or Class~II. But most of the SED models have no accreting 
envelope, suggesting stage~II.

\begin{figure}[htp]
\includegraphics[width=90mm,angle=0]{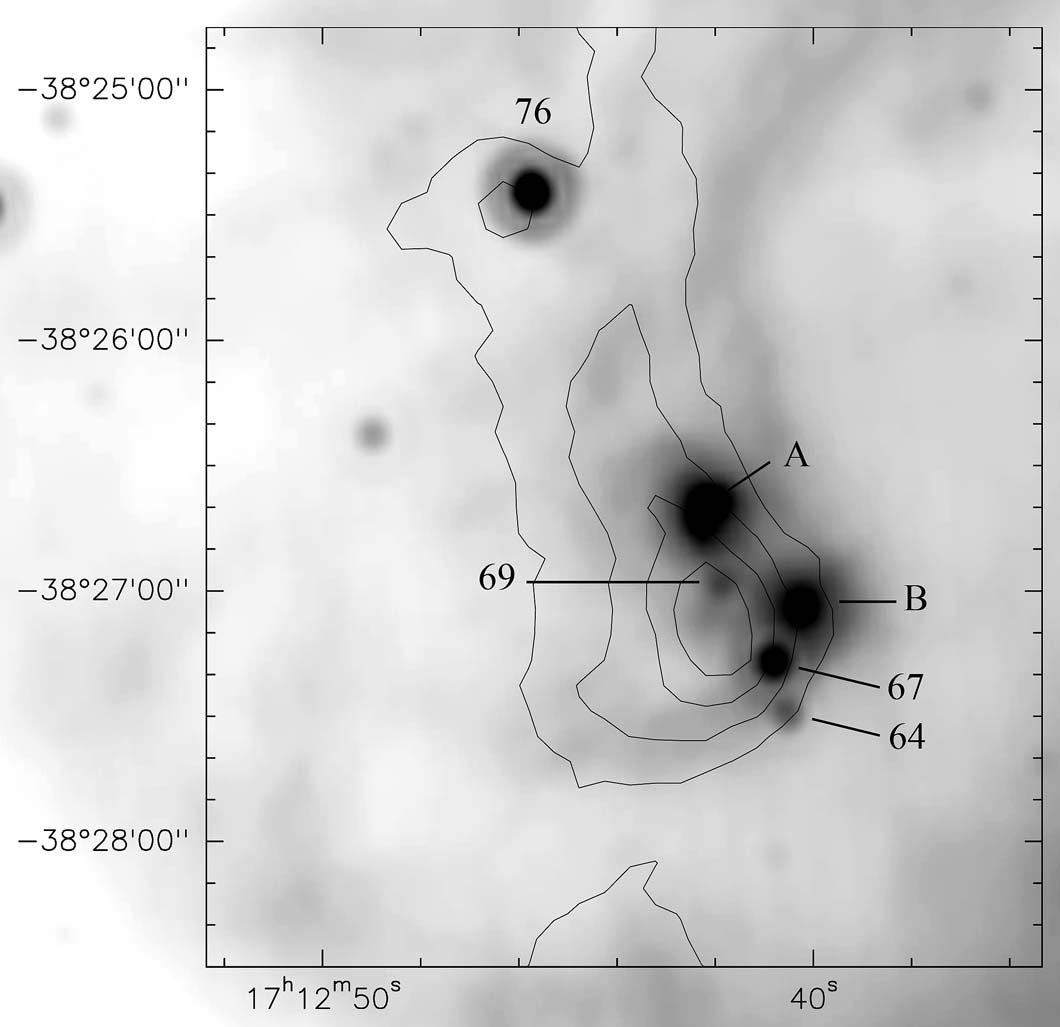}
\caption{Condensation 4: The background image is the Spitzer-MIPSGAL image 
at 24~$\mu$m; the contours show the 870~$\mu$m emission. 
The contour levels are 0.2~Jy/beam (used to estimate the mass of the 
condensation), 0.4, 0.6, and 0.8~Jy/beam.}
\label{cond4}
\end{figure}

$\bullet$ {\bf Condensation 5:} This is a group of 
structures emitting at 870~$\mu$m on the north-east 
border of RCW~120 (Fig.~\ref{cond5}). Its mass is in the range 
130--225~$M_{\sun}$. It contains several 
small spherical dust clumps in the direction of which YSOs are observed: 
clump a, of mass 15--26~$M_{\sun}$, possibly associated with \#107; 
clumps b1 + b2, of total mass 11--19~$M_{\sun}$, possibly associated 
with \#105 and \#103. Clump c, of mass 6--11~$M_{\sun}$, possibly 
associated with \#88, lies outside condensation~5.

Numerous 24~$\mu$m sources lie in the field of Fig.~\ref{cond5}. 
According to our indicators, sources \#88, \#91 and \#97 are Class~I YSOs;
sources \#103 and \#107 are probably intermediate between 
Class~I and Class~II; \#102 is a Class~II; and \#92 is intermediate between 
Class~II and Class~III. Magnitudes are missing for three sources; we have 
used the slope of the SED between 3.6~$\mu$m and 24~$\mu$m to classify them. 
Sources \#101 and \#105 have strongly rising SEDs (slopes   
1.9 and 1.6 respectively) characteristic of Class~I. Source \#114 is a Class~II source 
(decreasing slope of $-1.15$). The nature of \#113 is very uncertain, probably 
because it lies too close to the very bright \#114. 

We have used the Robitaille et al.\ SED fitting tool to model 
the SEDs of \#88 and \#105. The parameters of the best fit model are given 
in Table~\ref{models}. These two sources are rather similar, in an 
evolutionary stage~I, with a massive disk. We tried to model \#103 and \#107 
but the results are very uncertain (both stage~I and stage~II are obtained); 
far-IR measurements are clearly needed to constrain their SEDs.

\begin{figure}[htp]
\includegraphics[width=90mm,angle=0]{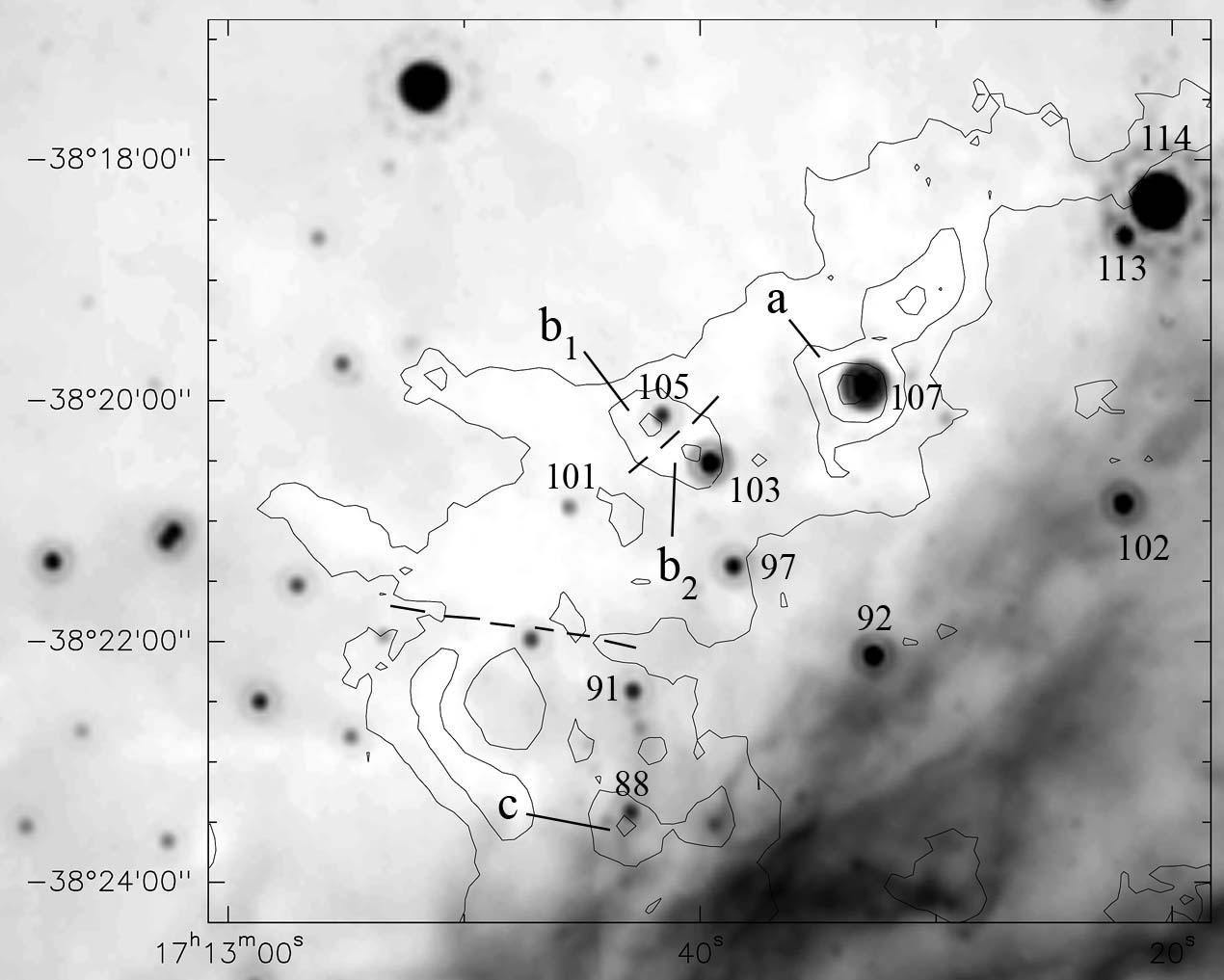}
\caption{Condensation 5: The background image is the Spitzer-MIPSGAL frame 
at 24~$\mu$m; the contours correspond to the 870~$\mu$m emission. 
The contour levels are 0.1~Jy/beam (used to estimate the mass of the 
condensation), 0.3, 0.5, and 0.7~Jy/beam; the dashed line shows
the adopted southern limit of the condensation.}
\label{cond5}
\end{figure}

$\bullet$ {\bf Condensation 6:} This lies on the north-west border of RCW~120, 
apparently far from the main IF. However a north-south 24~$\mu$m filament, 
probably tracing the emission of dust in the PDR, is seen adjacent to 
this condensation (Fig.~\ref{cond6}). 
Also, diffuse H$\alpha$ emission is observed nearby 
(Fig.~\ref{apexmap}). Thus this condensation is submitted to the 
pressure of the ionized gas. Its mass is in the range 
52--90~$M_{\sun}$; its peak column density corresponds to a visual 
extinction of 18\,mag. It appears as an IRDC on both the 8.0~$\mu$m and 
24~$\mu$m images.

Fig.~\ref{cond6} shows the 870~$\mu$m emission contours superimposed on 
the 24~$\mu$m image. Condensation~6 contains a small cluster of at least 
three YSOs, separated by $\sim$0.17~pc (in projection). YSO \#106, 
very bright at 24~$\mu$m (possibly saturated), is also detected at 70~$\mu$m; 
it lies 0.08~pc away from the dust emission peak. All the indicators show  
that these three YSOs are Class~I (\#108) or flat spectrum 
sources (\#106, \#110).

We tried to model the SED of object \#106 using the Robitaille 
et al.\ fitting tool. The evolutionary stage of this source is uncertain: 
the best model, with a central object of mass $2~M_{\sun}$ and an 
accreting envelope, corresponds to stage~I (Table~\ref{models}), but other 
good models have more massive central objects and no accreting envelopes,  
and thus correspond to stage~II sources. 

The other stars in the field of Fig.~\ref{cond6} are Class~II (\#99, \#100) or 
intermediate Class~II--Class~III (\#116). Fig.~\ref{cond6} also shows 
a small faint region of 24~$\mu$m extended emission which we call region~C. Its 
central star has 2MASS magnitudes ($J=12.70$, $H=11.99$, $K=11.65$) 
corresponding to a B9V star with a visual extinction of the order of 
6.1~mag. We suggest that the extended 24~$\mu$m emission 
comes from a PDR excited by this star. Region C and its exciting star are very 
similar to regions A and B in condensation 4, which we believe to be 
second-generation objects.

\begin{figure}[htp]
\includegraphics[width=90mm,angle=0]{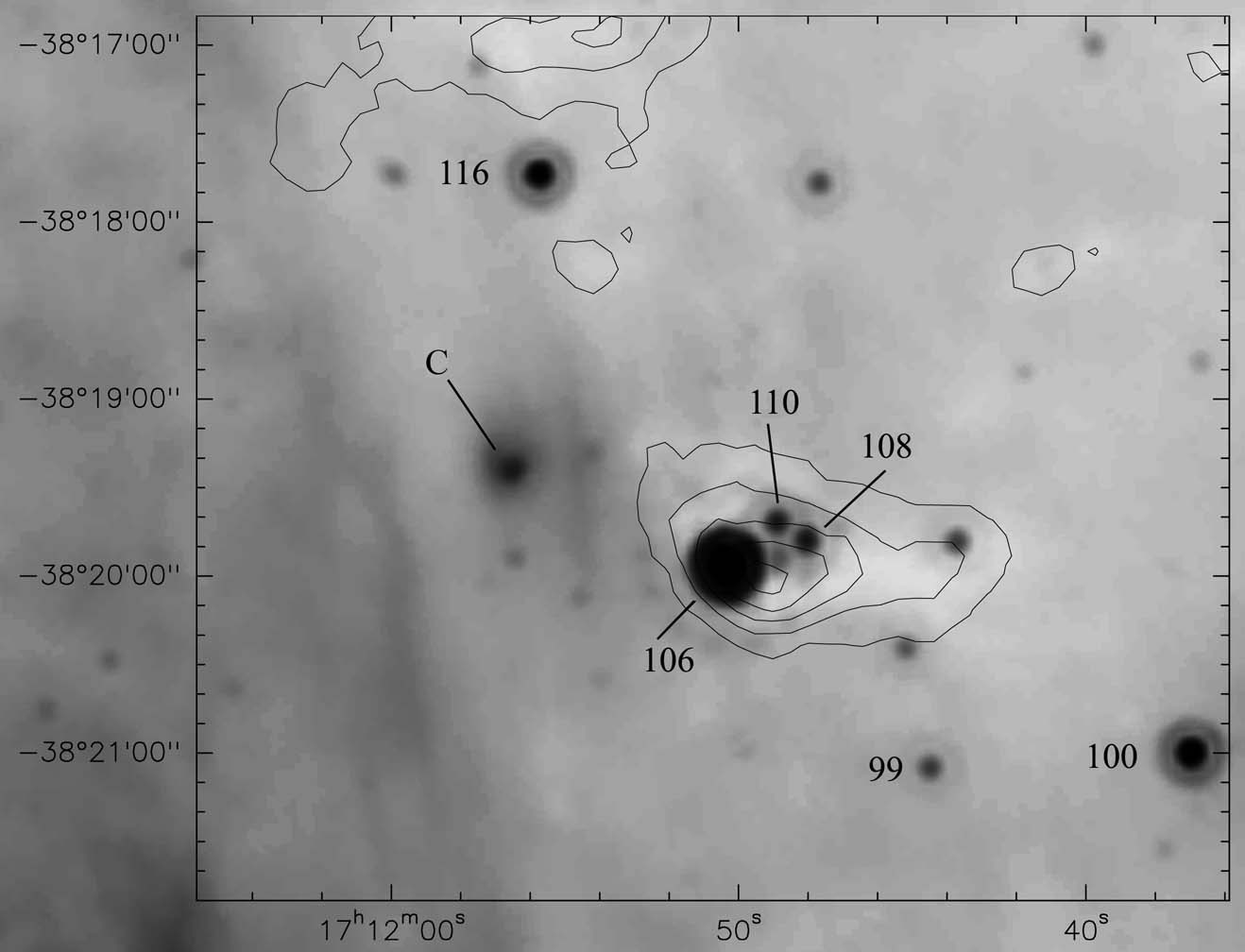}
\caption{Condensation 6 and its surrounding. The background image is 
the Spitzer-MIPSGAL frame at 24~$\mu$m. The contours correspond to the 
870~$\mu$m emission. The contour levels are 0.2~Jy/beam 
(used to estimate the mass of the condensation), 0.4, 0.6, 
0.8, and 1.0~Jy/beam.}
\label{cond6}
\end{figure}

%
\begin{table*}
\caption{Parameters of the ``best fit models'' for the SEDs of a few sources, 
obtained using the SED fitting tool of Robitaille et al.~(\cite{rob07})}
\begin{tabular}{c l r c  c c c}
  \hline\hline
  YSO & $M_{\rm star}$ & $T_{\rm star}$ & $M_{\rm disk}$ & 
   $\dot{M}_{\rm envelope}$ & $L$ & comment \\
     & ($M_{\sun}$) & (K)  & ($M_{\sun}$)   
  & ($M_{\sun}~yr^{-1}$) & ($L_{\sun}$)  & \\
  \hline
        &            &      &     &  &   &    \\
\#38 & 3.2 & 4350 & 2.4 10$^{-3}$ (10$^{-3}$--10$^{-1}$) & 1.7 10$^{-4}$ & 98 & stage~I \\
\#61 & 0.3 & 3370 & 4.3 10$^{-2}$ & 7.3 10$^{-5}$ & 20 & stage~I \\
\#51 & 1.3  & 4355  & 6.3 10$^{-2}$ & 1.8 10$^{-6}$ (10$^{-6}$--10$^{-4}$) &  17 & stage~I-II \\
\#59 & 6.6  & 19700 & 1.0 10$^{-5}$ (10$^{-5}$--10$^{-4}$) & 0      & 1320 & stage~II-III \\
\#67 & 5.1 & 7554  & 1.7 10$^{-3}$ (10$^{-5}$--10$^{-1}$) & 2.2 10$^{-7}$ & 280 & stage~II \\
\#88 & 4.1 & 4400 & 4.4 10$^{-2}$ (10$^{-3}$--10$^{-1}$) & 3 10$^{-4}$ & 130 & stage~I \\
\#105 & 1.7 & 4200 & 4.5 10$^{-2}$ (10$^{-3}$--10$^{-2}$) & 6 10$^{-4}$ & 38 & stage~I \\
\#106 & 2.0 & 4160  & 2.210$^{-1}$ & 1.2 10$^{-5}$ & 105 & uncertain\\
\hline
  \label{models}
\end{tabular}
\end{table*}

\subsection{The structure of the PDR and the long-distance influence of the 
\HII\ region}

At first glance RCW~120 appears to be a perfect spherical bubble. However, as 
stressed by ZA07, the ionized gas is trying to flow out of the bubble, and
consequently the 
\HII\ region is elongated towards the north, where the surrounding 
interstellar medium is of lower density. This morphology is 
delineated by the shape of the ionization front and by that of the low 
intensity 870~$\mu$m emission zones, as shown in Fig.~\ref{apexM1}, 
a composite image of the 870~$\mu$m emission in turquoise 
and of an unsharp-masked image at 24~$\mu$m in red. The unsharp masking 
emphasizes the small-scale structures of the 24~$\mu$m emission, both 
in the ionized gas and in the PDR. The black arrows at the top of 
Fig.~\ref{apexM1} point to low intensity 870~$\mu$m emission 
zones which close the northern extension of the ionized bubble; these zones
are associated with 24~$\mu$m filaments in the adjacent PDR. 
The upper arrow shows that a low intensity 870~$\mu$m emission region joins 
condensation 5 and condensation 9. Here again we are observing the layer 
of neutral material collected around the ionized region. 

Fig.~\ref{haM1} shows the extent of the PDR. This is a colour composite image of 
the H$\alpha$ emission, in red, and of the unsharp-masked image 
at 24~$\mu$m in turquoise. We see in red the H$\alpha$ emission of the 
ionized gas inside the bubble, and in pink a zone of diffuse low intensity 
H$\alpha$ emission from the PDR surrounding the bubble. This ionized PDR 
is extended. The unsharp-masked image at 24~$\mu$m shows the structures of 
the dust emission from the PDR, covering about the same zone as the 
diffuse H$\alpha$ emission. The ionized PDR extends over an area with about 
twice the radius of the central
\HII\ region. As stressed by ZA07, the IF is porous, and the ionizing radiation 
escapes to the outside through 
small holes. This ionized PDR is probably of low density, but 
its existence demonstrates the long-distance influence of the \HII\ region and 
its central exciting star. 

The 870~$\mu$m emission map shows radial features  
inside condensation 5 and also in the southern parts of RCW~120; there 
two bubbles are observed between condensations 1 and 2 (see the  
black arrows at the bottom of Fig.~\ref{apexM1}). 
Fig.~\ref{extinctionNE} shows, at 8.0~$\mu$m and with a high resolution, the 
radial structures inside condensation 5; only the pressure of the 
ionized gas seems able to form filaments perpendicular to the IF.

Thus star formation can probably be triggered  
far from the ionization front -- for example in condensation~6, which 
is in contact with the extended PDR.

\begin{figure}[htp]
\includegraphics[width=90mm,angle=0]{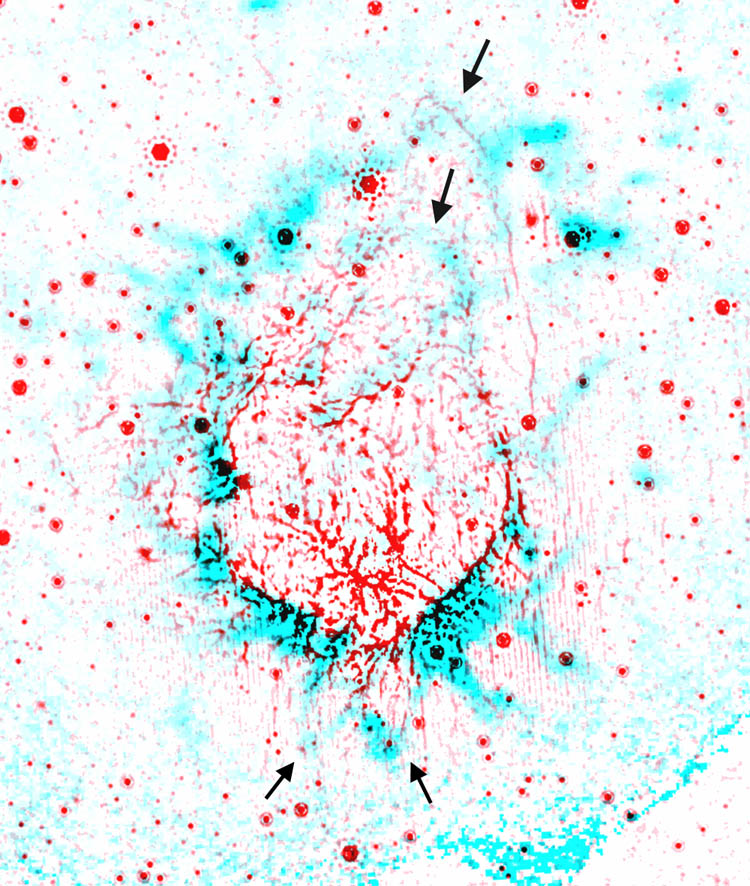}
\caption{Colour composite image, with the 870~$\mu$m emission in turquoise, 
and in red an unsharp-masked image at 24~$\mu$m which emphasizes 
the small-scale low intensity structures of the dust emission from both 
the ionized region and the surrounding PDR. The black arrows at the top 
of the figure point to the collected neutral material closing the northern 
extension of the ionized bubble. The arrows at the bottom of the figure 
point to the bubbles formed by the ionized gas flowing away from the 
central \HII\ region.}
\label{apexM1}
\end{figure}

\begin{figure}[htp]
\includegraphics[width=90mm]{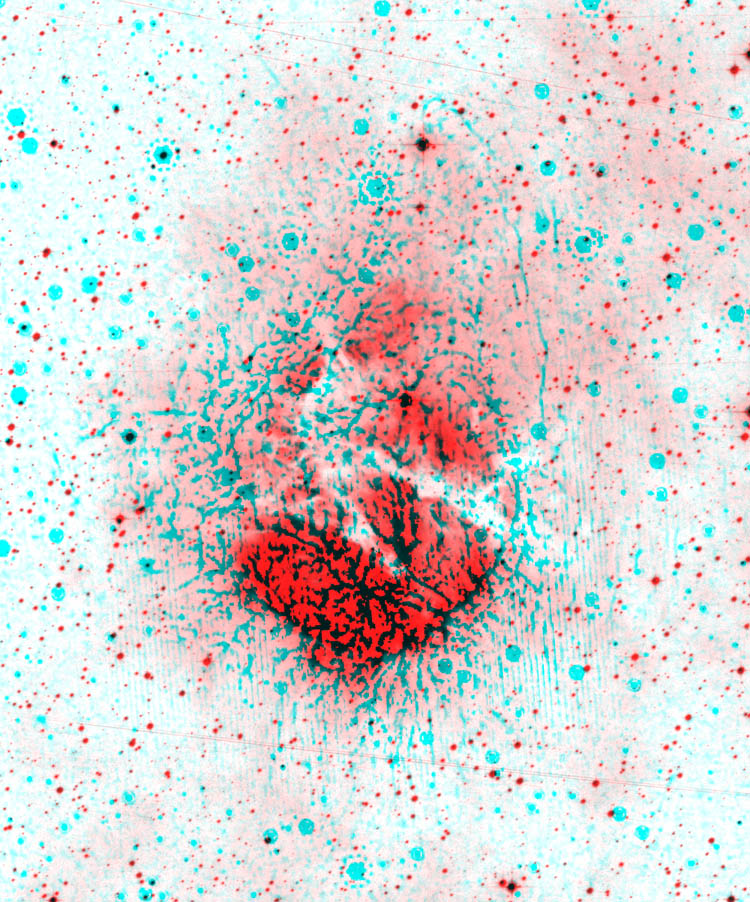}
\caption{Colour composite image, with the H$\alpha$ emission of the ionized 
gas in red, and the unsharp-masked image at 24~$\mu$m in turquoise.}
\label{haM1}
\end{figure}

\subsection{An unanswered question}

A question raised by ZA07 remains unanswered: what is the three-dimensional  
geometry of the RCW~120 bubble? Is there a preferential plane for 
the accumulation of neutral material during the expansion of RCW~120? 
We have identified only two Class~I sources in the direction of the 
ionized gas (\#61 and \#65), whereas more should be observed if star 
formation is at work in a spherical collected shell. According to 
the simulations of Krumholtz et al.~(\cite{kru07}), the presence 
of a magnetic field inhibits motion of the gas across 
magnetic field lines, leading to a concentration of collected material in a 
preferential plane. This point, important for the study of star formation 
triggered by \HII\ regions, will possibly be addressed via submillimetre  
polarimetric observations.

\section{Conclusions}

In this paper we present new observations, obtained with the 
APEX-LABOCA camera, of the thermal dust emission at 870~$\mu$m. 
The higher angular resolution and signal-to-noise ratio 
of these observations, compared to the previous 1.2-mm observations, 
allow the detection of low brightness structures and permit more accurate
mass estimates. The observations confirm 
the presence of a shell of dense neutral material collected around the 
ionized region during its expansion. This shell is now observed 
all around RCW~120, even at the border of its northern extension. 
The shell's mass is greater than 2000~$M_{\sun}$; we confirm that 
it is fragmented, with massive fragments of a few tens or hundreds of 
solar masses. We also confirm that a very massive ($\sim 250~M_{\sun}$) 
and compact (0.095~pc~$\times$~0.05~pc) core lies adjacent to the 
ionization front in the most massive fragment.

We present and discuss mid-IR observations at 24~$\mu$m 
and 70~$\mu$m, obtained by Spitzer-MIPSGAL. We have measured 138 sources 
at 24~$\mu$m and six at 70~$\mu$m. These observations provide new 
indicators to better distinguish between stars and YSOs, and to better
distinguish between the evolutionary 
stages (Class~I or Class~II) of the sources. Robitaille et al.~(\cite{rob06})  
have shown the uncertainties tainting these classifications. RCW~120 is 
a good illustration of the difficulty of this study. The main uncertainty
is due to the objects' external extinction, which is not known; highly 
reddened stars can be mistaken for Class~II sources, and highly reddened 
Class~II sources for Class~I sources. We have used the 870~$\mu$m 
emission to determine the maximum external extinction which may 
affect a given 24~$\mu$m source.

We have detected 21 Class~I sources, 18 sources possibly intermediate 
between Class~I and Class~II, 29 Class~II sources, and nine more uncertain 
Class~II--Class~III sources. A large fraction of the Class~I or 
intermediate Class~I--Class~II sources are observed in the direction 
of the shell of collected neutral material surrounding the ionized region. 
This distribution demonstrates that their formation has been triggered 
by the expanding \HII\ region. 

As stressed by ZA07, we believe that several 
different mechanisms are at work simultaneously to form stars in the 
collected layer. The mechanisms that we tentatively identify are 
small-scale gravitational instabilities in condensation~1, dynamical 
instabilities of the IF in condensation~2, and interaction of the IF with 
a pre-existing molecular condensation in condensation~4. The collect \& 
collapse process, which implies large-scale gravitational instabilities 
of the collected layer along its surface, is possibly at work in the 
dense core of condensation~1, which harbours a source detected 
only at 70~$\mu$m. High-resolution kinematic observations are 
necessary to ensure that a Class~0 source is present there, building 
its mass by accreting  material in the dense core; enough mass 
is present there ($\sim\,250$~$M_{\sun}$) to allow the formation 
of a massive star.

We have also detected, at 24~$\mu$m, in the direction of 
condensation~1, a very unusual and remarkable feature: a 
chain of about eleven Class~I or flat spectrum sources, 
parallel to the IF. These YSOs have approximately solar masses, and 
are regularly spaced by some 0.1~pc. To our knowledge it is one of 
the best illustrations ever found, in the domain of star formation, 
of the Jeans gravitational instability at work.

Finally, we have shown that the \HII\ region, via the UV radiation of its 
central exciting star, has a long distance influence upon its 
surrounding medium. If $R$ is the radius of the \HII\ region,  
the low density ionized PDR, and the associated 
dust emitting at 24~$\mu$m, extends a further $R$ outward from the IF. 
This influence is also 
reflected by the shape of the collected layer, which shows radial 
features and bubbles. We conclude that the \HII\ region can probably 
trigger star formation far from the IF. This remains to be 
confirmed by other examples of \HII\ regions.

\begin{acknowledgements}

We thank the APEX staff for their support during the observations 
with LABOCA, and A. Beelen who performed the first reduction 
of the data. We thank the anonymous referee for his suggestions and comments. 
This research has made use of the SIMBAD data base, 
operated at CDS, Strasbourg, France, and of the interactive sky
atlas Aladin (Bonnarel et al.\ \cite{bon00}). This work is based in part 
on observations made with the Spitzer Space Telescope, which is operated 
by the Jet Propulsion Laboratory, California
Institute of Technology, under contract with NASA. We have made use
of the NASA/IPAC Infrared Science Archive to obtain data products from 
the 2MASS, Spitzer-GLIMPSE and Spitzer-MIPS surveys.

\end{acknowledgements}




\begin{thebibliography}{}

\bibitem[2004]{all04} Allen, L.E., Calvet, N., D'Alessio, P., et al. 2004, ApJS, 154, 363
\bibitem[1984]{ave84} Avedisova, V.S., \& Kondratenko, G.I. 1984, Nauchnye Informatsii, 56, 59
\bibitem[2003]{ben03} Benjamin, R.A., Churchwell, E., Babler, B.L., et al. 2003, PASP, 115, 953
\bibitem[1978]{boh78} Bohlin, R.C., Savage, B.D., \& Drake, J.F. 1978, ApJ, 224, 132
\bibitem[2000]{bon00} Bonnarel, F., Fernique, P., Bienayme, O., et al. 2000, A\&ASS, 143, 33
\bibitem[2005]{car05} Carey, S.J., Noriega-Crespo, A., Price, S.D., et al. 2005, BAAS, 37, 1252
\bibitem[2006]{chu06} Churchwell, E., Povich, M. S., Allen, D., et al. 2006, ApJ, 649, 759
\bibitem[2008]{deh08} Deharveng, L., Zavagno, A. 2008, Handbook of Star Forming Regions, Vol. II, B. Reipurth, ed.
\bibitem[1996]{gar96} Garc\'{\i}a-Segura, G., \& Franco, J. 1996, ApJ, 469, 171
\bibitem[2005]{hen95} Henning, Th., Michel, B., Stognienko, R. 1995, P\&SS, 43, 1333
\bibitem[1983]{hil83} Hildebrand, R.H. 1983 Q.Jl. R. astr. Soc., 24, 267
\bibitem[1982]{hou82} Houk, N. 1982, ({\it Michigan Catalogue of Two-dimensional Spectral Types for the HD stars}), Vol.~3 (Ann Arbor: Univ. Michigan)
\bibitem[2005]{ind05} Indebetouw, R., Mathis, J.S., Babler, B.L., et al. 2005, ApJ, 619, 931
\bibitem[2007]{ind07} Indebetouw, R., Robitaille, T.P., Whitney, B.A., et al. 2007, ApJ, 666, 321
\bibitem[2006]{joh06} Johnstone, D., Matthews, H., Mitchell, G.F. 2006, ApJ, 639, 259
\bibitem[2003]{kes03} Kessel-Deynet, O., Burkert, A. 2003, MNRAS, 338, 545
\bibitem[2006]{kir06} Kirk, H., Johnstone, D., Di Franchesco, J. 2006, ApJ, 646, 1009
\bibitem[2007]{kru07} Krumholz, M.R., Stone, J.M., Gardiner, T.A. 2007, ApJ, 671, 518
\bibitem[1992]{lad92} Lada, C.J., \& Adams, F.C. 1992, ApJ, 393, 278
\bibitem[1994]{lef94} Lefloch, B., Lazareff, B. 1994, A\&A, 289, 559
\bibitem[2006]{mar06} Martins, F., Plez, B. 2006, A\&A, 457, 637
\bibitem[2003]{mot03} Motte, F., Schilke, P., Lis, D.C. 2003, ApJ,582, 277
\bibitem[1994]{oss94} Ossenkopf, V., Henning, T. 1994 A\&A, 291, 943
\bibitem[2005]{par05} Parker, Q.A., Phillipps, S., Pierce, M. J., et al. 2005, MNRAS, 362, 689
\bibitem[2007]{reb07} Rebull, L.M., Stapelfeldt, K.R., Evans II, N.J., et al. 2007, ApJSS, 171, 447
\bibitem[1985]{rie85} Rieke, G.\ H., \& Lebofsky, M.\ J.\ 1985, ApJ, 288, 618
\bibitem[2006]{rob06} Robitaille, T.P., Whitney, B.A., Indebetouw, R., \& Wood, K.,  
Denzmore, P. 2006, ApJS, 167, 256
\bibitem[2007]{rob07} Robitaille, T.P., Whitney, B.A., Indebetouw, R., \& Wood, K. 
2007, ApJS, 169, 328
\bibitem[2008]{rob08} Robitaille, T.P., Meade, M.R., Babler, B.L. et al. 2008, astro-ph/0809.1654v1
\bibitem[2007]{sir07} Siringo, G., Weiss, A., Kreysa, E., et al. 2007, The Messenger, 129, 2
\bibitem[1987]{ste87} Stetson, P.B. 1987, PASP, 99, 191
\bibitem[2002]{sug02} Sugitani, K., Tamura, M., Nakajima, Y., et al. 2002, ApJ, 565, L25
\bibitem[2000]{tok00}  Tokunaga, A.T. 2000, Allen's Astrophysical Quantities, 4th edition,
ed. A.N. Cox, Springer-Verlag (New York), p. 143
\bibitem[2007]{vig07} Vig, S., Testi, L., Walmsley, M., et al. 2007, A\&A, 470, 977
\bibitem[1999]{whi99} White, G.J., Nelson, R.P., Holland, W.S., et al. 1999, A\&A, 342, 233
\bibitem[2003a]{whi03a} Whitney, B.A., Wood, K., Bjorkman, J.E., Wolff, M.J. 2003a, ApJ, 591, 1049
\bibitem[2003b]{whi03b} Whitney, B.A., Wood, K., Bjorkman, J.E., Cohen, M. 2003b, ApJ, 598, 1079
\bibitem[2004]{whi04} Whitney, B.A., Indebetouw, R., Bjorkman, J., Wood, K. 2004, ApJ, 617, 1177
\bibitem[2003]{you03} Young, C.H., Shirley, Y.L., Evans II, N.J. 2003, ApJSS, 145, 111
\bibitem[2007]{zav07} Zavagno, A., Pomar\`es, M., Deharveng, L., et al. 2007, A\&A, 472, 835 (ZA07)
\end{thebibliography}
\end{document}